\documentclass[11pt]{article} 

\usepackage[top=1in,bottom=1in,left=1in,right=1in]{geometry} 
\RequirePackage{amsthm,amsmath,amsfonts,amssymb}
\RequirePackage[numbers]{natbib}
\RequirePackage[colorlinks,allcolors=blue,urlcolor=blue]{hyperref}
\RequirePackage{graphicx}

\usepackage{color, amsmath, amssymb, amsbsy, amsthm, graphicx, bbm, amsfonts, bm, dsfont, courier, subfig}
\usepackage[toc,page]{appendix}
\usepackage{booktabs}
\usepackage[flushleft]{threeparttable}
\usepackage{latexsym}
\usepackage[dvipsnames]{xcolor}
\usepackage{comment}
\usepackage{wrapfig}
\usepackage{graphicx}
\usepackage{graphics}
\usepackage{algorithm}
\usepackage{algorithmic}
\usepackage{setspace}

\usepackage{soul}
\usepackage{threeparttable,booktabs}
\usepackage{etoolbox}
\appto\TPTnoteSettings{\footnotesize}
\newtheoremstyle{theoremstyle}
{\topsep} % Space above
{\topsep} % Space below
{\itshape} % Body font
{} % Indent amount
{} % Theorem head font
{} % Punctuation after theorem head
{.5em} % Space after theorem head
{\color{black}\ifthenelse{\equal{#3}{}}{{\bfseries #1 #2}}{{\bfseries #1 #2 (#3)}}}
\newtheoremstyle{theoremstylealt}
{\topsep} % Space above
{\topsep} % Space below
{\itshape} % Body font
{} % Indent amount
{} % Theorem head font
{} % Punctuation after theorem head
{.5em} % Space after theorem head
{\color{black}\ifthenelse{\equal{#3}{}}{{\bfseries #1 #2$^\prime$}}{{\bfseries #1 #2$^\prime$ (#3)}}}
\newtheoremstyle{examplestyle}
{\topsep} % Space above
{\topsep} % Space below
{} % Body font
{} % Indent amount
{} % Theorem head font
{} % Punctuation after theorem head
{.5em} % Space after theorem head
{\color{black}\ifthenelse{\equal{#3}{}}{{\bfseries #1 #2}}{{\bfseries #1 #2 (#3)}}}
\theoremstyle{theoremstyle}\newtheorem{thm}{Theorem}
\theoremstyle{theoremstylealt}
\theoremstyle{theoremstyle}     
\theoremstyle{theoremstyle}\newtheorem{lem}{Lemma}  
\theoremstyle{theoremstyle}\newtheorem{coro}{Corollary}        
\theoremstyle{theoremstyle}
\theoremstyle{theoremstyle}\newtheorem{assumption}{Assumption}
\theoremstyle{theoremstylealt}\newtheorem{assumptionalt}{Assumption}
\theoremstyle{theoremstyle}\newtheorem{proc}{Step}

\theoremstyle{theoremstyle}

\theoremstyle{theoremstyle}

\theoremstyle{examplestyle}
\theoremstyle{examplestyle}
\theoremstyle{examplestyle}

\newcommand{\Expectation}{\mathbb{E}}

\newcommand{\Var}{\mathbb{V}}

\newcommand{\Prob}{\mathbb{P}}
\newcommand{\Indicator}{\mathbf{1}}

\def\independenT#1#2{\mathrel{\rlap{$#1#2$}\mkern2mu{#1#2}}}

\newcommand{\toProb}{\overset{\mathrm{p}}{\to}}

\newcommand{\precsimProb}{\precsim_{\mathrm{p}}}
\newcommand{\precProb}{\prec_{\mathrm{p}}}

\newcommand{\asympProb}{\asymp_{\mathrm{p}}}

\newcommand{\diff}{\mathrm{d}}

\renewcommand{\epsilon}{\varepsilon}

\def \hat{\widehat}

\newcommand{\cw}[1]{{\color{black} #1}}

\newcommand{\xwm}[1]{{\color{black} #1}}

\newcommand\independent{\protect\mathpalette{\protect\independenT}{\perp}}
\def\independenT#1#2{\mathrel{\rlap{$#1#2$}\mkern2mu{#1#2}}}
\usepackage{array}
\newcolumntype{H}{>{\setbox0=\hbox\bgroup}c<{\egroup}@{}}

\begin{document}
\title{Breaking the Winner's Curse in Mendelian Randomization: Rerandomized Inverse Variance Weighted Estimator
\thanks{
The authors are in alphabetical order. We would like to thank the editor, the associate editor, and three anonymous reviewers for their comments and suggestions that significantly improved the paper. We also thank Matias D. Cattaneo, Wei Pan, Hongyu Zhao and  Meng Zhuo (alphabetically ordered) for their valuable feedback and thoughtful discussions. 
}}
\author{
Xinwei Ma\thanks{Department of Economics, University of California San Diego.} \and
Jingshen Wang\thanks{Division of Biostatistics, University of California Berkeley.} \and
Chong Wu\thanks{Department of Biostatistics, The University of Texas MD Anderson Cancer Center.}
} 
\maketitle
		
		\begin{abstract}
		Developments in genome-wide association studies and the increasing availability of summary genetic association data have made the application of two-sample Mendelian Randomization (MR) with summary data increasingly popular. Conventional two-sample MR methods often employ the same sample for selecting relevant genetic variants and for constructing final causal estimates. Such a practice often leads to biased causal effect estimates due to the well known ``winner's curse'' phenomenon. To address this fundamental challenge, we first examine its consequence on causal effect estimation both theoretically and empirically. We then propose a novel framework that systematically breaks the winner's curse, leading to unbiased association effect estimates for the selected genetic variants. Building upon the proposed framework, we introduce a novel rerandomized inverse variance weighted estimator that is consistent when selection and parameter estimation are conducted on the same sample. Under appropriate conditions, we show that the proposed RIVW estimator for the causal effect converges to a normal distribution asymptotically and its variance can be well estimated. We illustrate the finite-sample performance of our approach through Monte Carlo experiments and two empirical examples. \\ \bigskip 
		
		\end{abstract}
		
\clearpage
\onehalfspacing
%%%%%%%%%%%%%%%%%%%%%%%%%%%%%%%%%%%%%%%%%%%%%%%%%%%%%%%%%%%%%%%%%%%%%%%%%%%%%%%%%
% Introduction
%%%%%%%%%%%%%%%%%%%%%%%%%%%%%%%%%%%%%%%%%%%%%%%%%%%%%%%%%%%%%%%%%%%%%%%%%%%%%%%%%
\section{Introduction}\label{Section-introduction}

%--------------------------------------------------------------------------------
% Motivation
%--------------------------------------------------------------------------------	
\subsection{Motivation}
	
Mendelian Randomization (MR) has become a popular tool for causal inference in observational studies. Its popularity partially ascribes to a fundamental limitation of observational data, where  causation cannot be simply inferred from the association between an exposure and a disease due to the issue of unmeasured confounders \citep{smith2004mendelian}. Since genetic variants are randomly assorted from parents and are fixed at conception, they provide a source of exogenous variation in the exposure. MR mitigates unmeasured confounding bias by incorporating these genetic variants as instrumental variables (IVs) \citep{lawlor2008mendelian}. In MR, the most commonly used genetic variants are single nucleotide polymorphisms (SNPs). 
	
The proliferation of genome-wide associate study (GWAS) data not only improves the statistical power of MR analyses thanks to the increased sample sizes, but also facilitates efficient evaluations of causal effects without the need to share information at individual level \citep{burgess2015using}. Two-sample MR, a design strategy where genetic associations with the exposure and the outcome are taken from separate samples in the same population, yields valid causal estimates even if concomitant data on the SNPs, exposure, and outcome are not available for any individual, provided that each included SNP is a valid IV \citep{burgess2013mendelian,pierce2013efficient}. To be precise, the included SNPs must be (i) associated with the exposure (relevance assumption), (ii) not associated with any unmeasured confounder (effective random assignment), and (iii) only associated with the outcome through the exposure (exclusion restriction). 
	
In practice, to make these assumptions plausible in the summary data setting where the whole-genome SNPs (about 10 million common variants) are harvested from GWAS, pre-selection procedures need to be conducted before carrying out the MR analysis. In particular, because the strength of each SNP can be tested empirically, practitioners routinely select SNPs that are strongly associated with the exposure to meet the relevance assumption. As it is well recognized in the literature \citep{gkatzionis2019contextualizing,lawlor2016commentary, sadreev2021navigating}, however, this pre-selection step can lead to biased causal effect estimates due to the well-known ``winner's curse'' phenomenon. This winner’s curse bias, which is the result of using the same sample to select the SNPs and to form the final causal estimate, is also borne out in our simulation and empirical studies (see Tables \ref{Table:simRes} and \ref{realdata1}).  

%--------------------------------------------------------------------------------
% Our contributions
%--------------------------------------------------------------------------------	
\subsection{Our contributions}
	
Given the issues mentioned above, the primary contribution of this paper is a novel MR framework and a Rerandomized Inverse Variance Weighted (RIVW) estimator that break the winner's curse in the two-sample MR with summary data setting. In addition, we provide a systematic study of their theoretical properties along with comprehensive evaluations of their finite-sample performance in simulation studies and real data analyses. We further break down our contribution as follows:

On the statistical methodology side, we demonstrate that the winner's curse pushes the estimated effect of SNP on exposure away from zero, and such distortion leads to an under-estimation bias in the classical two-sample Inverse Variance Weighted (IVW) estimator (Section \ref{Sec:selection-bias}). To break the winner's curse, we introduce pseudo SNP-exposure associations into the selection step, and then use Rao-Blackwellization to recover the correct center of the effect of selected SNPs on exposure (Section \ref{Sec:RIVW}). Because Rao-Blackwellization \citep{blackwell1947conditional} produces an estimator that is optimal by the mean-squared-error criterion among unbiased estimators, we make the fullest possible use of data from each SNP without incurring any selection bias from data re-use in SNP selection and parameter estimation. Moreover, we propose a novel RIVW estimator that not only eliminates winner's curse bias but also removes measurement error bias in the classical IVW approach.
	
On the theoretical side, we first establish an asymptotic normality result for the proposed RIVW estimator (Theorem \ref{thm:asy normality}). In particular, this result shows that our RIVW estimator is correctly centered despite the fact that the same data is used for SNP selection and causal effect estimation. In other words, the RIVW estimator is immune to winner's curse bias as well as measurement error bias. As part of this endeavor, we rigorously show that the Rao-Blackwellized SNP-exposure association is unbiased after selection, and we further provide an unbiased estimate of its variance (Section \ref{Sec:RB-theorey}). The latter turns out to be crucial for correcting measurement error bias. Next, we propose a consistent standard error motivated by the regression interpretation of the RIVW estimator (Theorem \ref{thm:valid standard error}). Different from other constructions in the literature, which can take different forms depending on if balanced horizontal pleiotropy is present, our standard error remains the same in both cases, alleviating the burden of choosing what type of assumptions should be used for practitioners (Corollary \ref{coro:pleiotropy} and Section \ref{Sec:balanced horizontal pleiotropy}). Third, because our theoretical developments require bounding moments of distributions related to the truncated normal with a diverging truncation threshold, some of the theoretical results can be of independent interest (see the Supplementary Material \cite{Ma-Wang-Wu-2022-AOS-Supp} for details).

On the practical side, we illustrate the finite-sample performance of our approach through Monte Carlo experiments (Section \ref{Sec:simulation}) and two empirical examples (Section \ref{Sec:real-data}), demonstrating that the winner's curse leads to biased causal estimates and such a bias can be corrected by the RIVW estimator. In particular, when analyzing the causal association between body mass index (BMI) and COVID-19 illness, our approach, in accordance with existing literature, confirms that high BMI is a causal risk factor for COVID-19 severity. Moreover, as our approach removes the downward bias caused by the winner's curse, we conjecture that the effect of high BMI on COVID-19 severity (odds ratio: 1.60) is higher than what the literature currently believes. Effectively communicating this fact to the general public and persuading people with obesity to take extra precautions may reduce the number of hospitalized COVID-19 patients and save lives (Section \ref{Sec:covid}).
	
%--------------------------------------------------------------------------------
% Existing literature
%--------------------------------------------------------------------------------
\subsection{Existing literature}
	
The winner's curse is well-recognized in the Mendelian randomization and human genomics literature as a fundamental challenge. To avoid this issue, the most common practice is to use a third independent dataset for instrument selection, known as ``three-sample'' MR designs. While appealing, finding a third independent sample might be difficult in practice, as practitioners tend to use the largest GWAS data (meta-analyzed results from most available cohorts) to improve power, leaving fewer options for the third independent dataset. In two-sample MR analyses with summary data,  MR estimators that break the winner's curse with rigorous statistical guarantees have been lacking. 
	
Other than the issue of winner's curse bias, we also face the challenge that all GWAS associations are estimated, that is, the associations between the SNPs and the exposure/outcome are measured with errors. Such a measurement error issue is also known as weak instrument bias in the MR literature; see \cite{andrews2019weak} and \cite{sadreev2021navigating} for more discussions and additional references. Without accounting for this issue, classical MR estimators (such as the IVW estimator) are often biased downwards. Our proposed RIVW estimator formulated in \eqref{eq:RIVW} removes this measurement error bias, making it immune to both sources of biases in the two-sample MR with summary data framework. See \cite{bowden2019improving,ye2019debiased,zhao2020statistical} for other recent attempts to tackle measurement error bias. 

By correcting winner's curse bias and measurement error bias, the proposed RIVW estimator improves upon the classical IVW approach, where the latter often serves as a benchmark in MR analyses  \citep{burgess2019guidelines}. For instance, in clinical studies \citep[see][for example]{andrews2019association,larsson2017modifiable,leong2021cardiometabolic,ponsford2020cardiometabolic,siedlinski2020white,wu2021systematic}, researchers use causal effect estimates obtained via IVW as their main results and validate their findings through sensitivity and robust MR analyses. This practice highlights an urgent need to provide an unbiased causal effect estimator of the IVW-type, which is our focus here. In this paper, we also demonstrate that the proposed RIVW estimator is robust to the presence of balanced horizontal pleiotropy, and we leave general robust MR analyses to future research.
	
Some existing approaches use all SNPs for MR analyses (i.e., without instrument selection). For example, \cite{ye2019debiased} propose a debiased IVW estimator which incorporates all SNPs into the estimator to avoid winner's curse bias. Likelihood-based approaches \citep{zhao2019powerful}, which are also robust to measurement error bias, are arguably more complex than our current construction. Moreover, as mentioned by \cite{zhao2019powerful}, the likelihood-based estimator may not deliver unique estimates for certain data generating processes. Our RIVW estimator, on the other hand, is easy to construct and does not require any numerical optimization.  
		
%--------------------------------------------------------------------------------
% Two-sample MR with summary data framework
%--------------------------------------------------------------------------------
\subsection{Two-sample MR with summary data framework}
	
In this section, we introduce two-sample MR framework with summary data, whereby genetic associations with the exposure and outcome are gleaned from independent samples to furnish a ``two-sample'' analysis.  

Throughout this paper, we denote by $\beta$ the causal effect of an exposure variable $X$ on an outcome variable $Y$. In the presence of an unmeasured confounder $U$, Mendelian Randomization (MR) analyses incorporate $p$ genetic variants $G_1, \ldots, G_p$ as instrumental variables (IVs). Here, genetic variant $G_j\in\{0,1,2\}$ represents the number of alleles of a SNP $j$ assigned to an individual. Existing MR literature 
starts with the following model 
\begin{align}\label{eq:DGP-linear-model}
&Y = \beta X +  \theta_Y U + E_Y, \quad X = \sum_{j=1}^p \gamma_j G_j + \theta_X U + E_X, 
\end{align}
where $ \gamma_1,\ldots, \gamma_p, \beta, \theta_X , \theta_Y$ are unknown parameters. $E_X $ and $E_Y$ are mutually independent noise variables that are also independent with $\big( G_1, \ldots, G_p, U \big)$. We later discuss a more general model setup allowing for direct dependence between $Y$ and $G_j$ in Section \ref{Sec:balanced horizontal pleiotropy}.

\begin{table}
\begin{center}
\begin{tabular}{ccc}
\hline \hline
&  Individual-level data\phantom{aaa} & Publicly available summary data \\[2ex] 
1. \textit{Exposure dataset }& $\Big\{ \big(X_i^*, G_{ij}^*\big)\Big\}_{i=1}^{n_X}$ &  $\Big\{ \big(\hat{\gamma}_j, \sigma_{X_j}\big)\Big\}_{j=1}^p$  \\[2ex]
2. \textit{Outcome dataset} & 	$\Big\{ \big(Y_i, G_{ij}\big)\Big\}_{i=1}^{n_Y}$ &  $\Big\{ \big(\hat{\Gamma}_j, \sigma_{Y_j}\big)\Big\}_{j=1}^p$  \\
\hline \hline
\end{tabular}
\end{center}
\caption{\label{table:Two-sample-MR-table-illustration} Two sample MR with summary data illustration. $\hat{\gamma}_j$ is obtained by regressing $(X_1^*, \ldots, X_{n_X}^*)$ on $ (G_{1j}^*, \ldots, G_{n_X,j}^*)$, and $\hat{\Gamma}_j$ follows from the regression of $(Y_1, \ldots, Y_{n_Y})$ on $ (G_{1j}, \ldots, G_{n_Y,j})$. The corresponding standard errors are $\sigma_{X_j}$ and $\sigma_{Y_j}$, respectively. }
\end{table}
	
Because sharing individual level data is usually impractical due to privacy constraints \citep{burgess2015using}, two-sample MR with summary data attempts to conduct MR analyses using summarized estimates of SNP-exposure and SNP-outcome associations based on two large and independent datasets (see an illustration in Table \ref{table:Two-sample-MR-table-illustration}). To be more precise, we have access to the marginal regression coefficients and standard errors for all SNPs, $\{ (\hat{\gamma}_j, \sigma_{X_j})\}_{j=1}^p$, which are obtained by regressing the exposure on each SNP separately on the exposure dataset $\{ (X_i^*, G_{ij}^*)\}_{i=1}^{n_X}$. Similarly,  by regressing the outcome on each SNP separately on the outcome dataset $\{ (Y_i, G_{ij})\}_{i=1}^{n_Y}$, we obtain the marginal regression coefficients and the standard errors, $\{ (\hat{\Gamma}_j, \sigma_{Y_j})\}_{j=1}^p$. 

To facilitate discussion, we start with the following assumption. We shall demonstrate in Section \ref{Sec:balanced horizontal pleiotropy} that the proposed method remains valid under a more general balanced horizontal pleiotropy Assumption \ref{assump:Measurement-error-pleiotropy}$^\prime$ (see Corollary \ref{coro:pleiotropy}). In Section \ref{Sec:extension-correlated-IV}, we will further extend our proposed method to allow for correlated SNPs.

\begin{assumption}[\normalfont Measurement error model]\label{assump:Measurement-error}
(i) For any $j\neq j'$, the pairs, $(\hat\Gamma_j,\hat\gamma_j)$ and $(\hat\Gamma_{j'},\hat\gamma_{j'})$ are mutually independent.\\
(ii) For each $j$, 
\begin{align*}
\begin{bmatrix}
\hat\Gamma_j \\
\hat\gamma_j
\end{bmatrix} \sim \mathcal{N}\left(\begin{bmatrix}
\Gamma_j\\
\gamma_j
\end{bmatrix}\ ,\ \begin{bmatrix}
{\sigma}_{Y_j}^2 & 0 \\
0 & {\sigma}_{X_j}^2
\end{bmatrix}\right),
\end{align*}
In addition, there exists some $\nu\to 0$, such that $\{\sigma_{Y_j}/\nu,\sigma_{X_j}/\nu:1\leq j\leq p\}$ are bounded and bounded away from zero. 
\end{assumption}

The normality assumption is plausible and is widely adopted in the MR literature, because  $\hat{\Gamma}_j$ and $\hat{\gamma}_j$ are estimated from modern GWAS with hundreds of thousands of samples, making the normal approximation highly accurate. To meet with the mutual independence assumption, one can conduct linkage disequilibrium  (LD) pruning \citep{purcell2007plink}  on the exposure dataset to pick independent SNPs.\footnote{\xwm{We note that this is a special case of linear instrumental variable models, in the sense that marginal association estimates are assumed to be independent. As a result, methods tailored to MR analyses (including ours) cannot be directly applied to more general instrumental variables models such as those discussed in \cite{Benjamin2012,chao2005consistent,Crudu2021,hansen2014instrumental,mikusheva2020inference,Zhu2018}. 
} } Because classical LD pruning selects independent SNPs without using the GWAS summary statistics information, distribution of the selected SNP-exposure associations remains unchanged. In Section \ref{Sec:extension-correlated-IV}, we propose a revised sigma-based LD pruning method that improves the efficiency of causal effect estimation. Finally, to simplify the presentation, we assume that the estimation uncertainties (i.e., $\sigma_{Y_i}$ and $\sigma_{X_j}$) are of the same order. While it is possible to generalize this condition in our theoretical analysis, it will unavoidably lead to more cumbersome notation. 

%%%%%%%%%%%%%%%%%%%%%%%%%%%%%%%%%%%%%%%%%%%%%%%%%%%%%%%%%%%%%%%%%%%%%%%%%%%%%%%%%
% Winner's curs
%%%%%%%%%%%%%%%%%%%%%%%%%%%%%%%%%%%%%%%%%%%%%%%%%%%%%%%%%%%%%%%%%%%%%%%%%%%%%%%%%	
\section{Winner's curse}\label{Sec:selection-bias}

In this section, we first introduce winner's curse bias in the two-sample MR with summary data framework. We then demonstrate how the winner's curse leads to the violation of Assumption \ref{assump:Measurement-error}. Lastly, we show empirically that the popular IVW estimator underestimates the true causal effect due to winner's curse bias (and measurement error bias). 
     
In MR analyses, to meet the relevance assumption, practitioners  often select a set of genetic instruments (i.e., SNPs) that are strongly associated with the exposure based on hard-thresholding individual $z$-scores: 
\begin{align}\label{eq:hard-thresholding-selection}
\widetilde{\mathcal{S}}_\lambda = \left\{ j: \ \Big| \frac{\hat{\gamma}_j}{\sigma_{X_j}} \Big| > \lambda, \ j =1, \ldots, p \right\},
\end{align}
where $\lambda$ is a pre-specified cutoff value.\footnote{\label{footnote:1}The cutoff value, $\lambda$, is often chosen to be $\Phi^{-1}(1-\alpha/2)$, which is the $(1-\alpha/2)$th quantile of the standard normal distribution. A large cutoff value $\lambda$ is often desired so that the selected SNPs meet the relevance assumption (for example, corresponding to $\alpha = 5\times 10^{-8}$ or $5\times 10^{-5}$). } This exercise casts ``winner's curse'' into MR analyses. 
	
Heuristically, the winner's curse, as discussed here, is caused by the tendency of the estimates $\{\hat{\gamma}_j, j\in \widetilde{\mathcal{S}}_\lambda\}$ to be more extreme than the underlying true association effects $\{\gamma_j, j\in \widetilde{\mathcal{S}}_\lambda\}$. It can be further understood from the ``file drawer effect,'' coined in \cite{rosenthal1979file}, which refers to the observation that statistically insignificant results are much less likely to be published and hence they remain in file drawers. To put this classical problem in the MR context, $\hat{\gamma}_j$ will be employed in the construction of the final causal effect estimator only when it is statistically significant. As a result, the distribution of $\hat{\gamma}_j$ after surviving the selection step is truncated Gaussian. Because the selection step will distort the distribution of $\hat{\gamma}_j$, it leads to the violation of Assumption \ref{assump:Measurement-error}. Classical MR estimators (such as two-sample IVW) incorporating instrument selection may suffer from the winner's curse, rendering both estimation and statistical inference problematic. 
	 
Although there is a general awareness in the MR literature that data-driven instrument selection affects subsequent estimation and inference, it remains pervasive in practice to ignore the consequence of this selection step. This is primarily due to the fact that, in an attempt to increase sample size and statistical power, most GWAS tend to meta-analyze results from  all available cohorts, making it challenging to find an independent (third) sample from the same population for instrument selection. To demonstrate the consequence of performing statistical inference with the same dataset for instrument selection, we examine in a simulation study the performance of IVW estimators. The IVW estimator employs summary associations obtained from two independent GWAS with SNP selection, and it estimates $\beta$ by a weighted regression of $\hat{\Gamma}_j$ on $\hat{\gamma}_j$: 
\begin{align}\label{eq:IVW-classical}
\hat{\beta}_{\lambda, \mathtt{IVW}} = \frac{ \sum_{j\in \widetilde{\mathcal{S}}_\lambda}\hat{\Gamma}_j \hat{\gamma}_j/\sigma_{Y_j}^2   }{ \sum_{j\in \widetilde{\mathcal{S}}_\lambda}\hat{\gamma}_j^2/\sigma_{Y_j}^2}.
\end{align}

\begin{figure}
\centering
\includegraphics[width=0.5\linewidth]{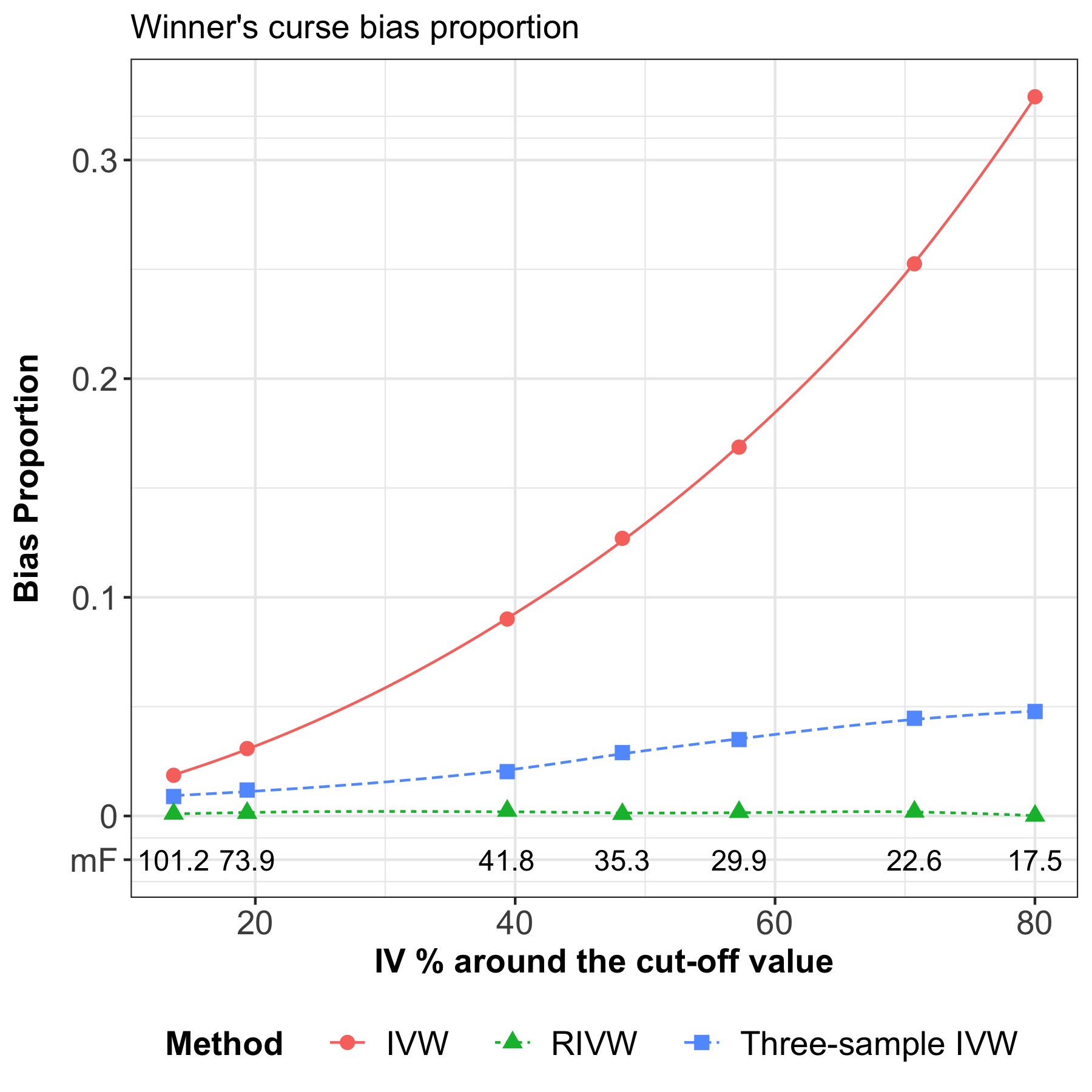}
\caption{Illustration of winner’s curse and measurement error bias. The IV proportion (x-axis) is calculated as the number of IVs with p-values lying between $5\times 10^{-8}$ and $5\times 10^{-10}$ divided by the number of selected IVs with p-value $<5\times 10^{-8}$. \cw{mF is mean F-statistic for the instruments used in three-sample IVW.}}	\label{fig:selectioin-bias}
\end{figure}

In Figure \ref{fig:selectioin-bias}, we showcase the performance of (i) the three-sample IVW estimator as a benchmark, (ii) the two-sample IVW estimator with instrument selection, and (iii) our proposed RIVW (to be discussed in Section \ref{Sec:RIVW}), where we report the bias proportion (absolute of Monte Carlo bias divided by the true parameter $\beta=0.2$) of each estimator. For comparison, we vary the number of IVs around the cutoff value $\lambda$, as this will affect the magnitude of winner's curse bias. To conserve space, we leave the details of our simulation design to Section \ref{Section:simulation-no-pleiotropy}.\footnote{we vary $\varepsilon_x^2 = \tau^2$ in the set $\{2\times 10^{-5}, 3\times 10^{-5}, 5\times 10^{-5},1\times 10^{-4}, 3\times 10^{-4}, 5\times 10^{-4}\}$, and $\pi_x = \pi_y$ in the set $\{0.005, 0.05\}$. Some combinations are ignored, because we require the heritability to be between 0 and 1, and that at least three IVs are selected for IVW with commonly used threshold $5\times 10^{-8}$. We set $\lambda=5\times 10^{-8}$ for the IVW estimator $\hat{\beta}_{\lambda, \mathtt{IVW}}$. To generate the SNP-exposure effects $\hat{\gamma}_j$ and SNP-outcome effects $\hat{\Gamma}_j$, we set the sample size  \cw{$n_X = n_Y = 100,000$}. The standard deviations are set to be $\sigma_{X_j} = 1/\sqrt{n_X}$ and $\sigma_{Y_j} = 1/\sqrt{n_Y}$. For three-sample IVW, we use a third independent exposure GWAS data with sample size $100,000$ for instrumental variable selection. }	

Results in Figure \ref{fig:selectioin-bias} demonstrate that the bias of $\hat{\beta}_{\lambda, \mathtt{IVW}} $ is larger if more SNP-exposure associations are around the selection cutoff $\lambda$. On the other hand, our proposed RIVW estimator is not sensitive to the change of such proportions, meaning that it is immune to winner's curse bias. Not surprisingly, the three-sample IVW estimator also does not suffer from winner's curse bias, although, as we mentioned earlier, finding a large third independent sample from the same population for instrument selection might be difficult in practice. In addition, we still observe a small residual bias for the three-sample IVW estimator. This is caused by measurement error bias. \xwm{We also provide mean F-statistic (mF) for each simulation setting in Figure \ref{fig:selectioin-bias}. Because instrument selection is employed by all three procedures, observed mean F-statistic is reasonably large. As a result, we only see moderate measurement error bias.} \xwm{We discuss in Section \ref{Section:measurement-error-bias} methods for correcting measurement error bias.}
	
%%%%%%%%%%%%%%%%%%%%%%%%%%%%%%%%%%%%%%%%%%%%%%%%%%%%%%%%%%%%%%%%%%%%%%%%%%%%%%%%%
% Rerandomized IVW estimator
%%%%%%%%%%%%%%%%%%%%%%%%%%%%%%%%%%%%%%%%%%%%%%%%%%%%%%%%%%%%%%%%%%%%%%%%%%%%%%%%%
\section{Rerandomized IVW estimator}\label{Sec:RIVW}
	
In this section, we first discuss how we lift the winner's curse after instrument selection and the intuition behind our method. We then introduce our rerandomized IVW (RIVW) estimator for estimating the true causal effect $\beta$. To streamline the presentation, theoretical results and their formal statements are presented in Section \ref{Sec:theory}. Section \ref{Sec:Extensions} considers several extensions to our framework, including (i) allowing for balanced horizontal pleiotropy, (ii) a novel smoothed RIVW estimator constructed via re-weighting the SNP-exposure associations, and (iii) an attempt to incorporate correlated IVs. 
	
%--------------------------------------------------------------------------------
% Winner's curse removal
%--------------------------------------------------------------------------------
\subsection{Winner's curse removal}\label{Sec:winner-curse-removal}
	
The fact that having a third independent sample for instrument selection lifts the winner's curse has a direct implication. \xwm{Recall that $\hat{\gamma}_j$ is the measured SNP-exposure association in the exposure dataset, and suppose $\hat{\gamma}_j'$ is another estimate of the SNP-exposure association (with standard error $\sigma_{X_j}'$) obtained from a third independent GWAS. Then the three-sample MR estimator is free of winner's curse bias, because the instrument selection event is \textit{independent} of the SNP-exposure association estimation, that is,}
\begin{align*}
\hat\gamma_{j}  \independent \underbrace{ \Big| \frac{\hat\gamma'_{j} }{\sigma'_{X_j}} \Big| > \lambda}_{\substack{\text{IV selection event} \\ \text{on a third GWAS}}}
\ \quad  \Longrightarrow \quad \mathbb{E}\Big[\hat\gamma_{j}   \ \Big|\  | \frac{\hat\gamma'_{j} }{\sigma'_{X_j}} | > \lambda\Big]  = \mathbb{E}[\hat\gamma_{j}] = \gamma_{j}  . 
\end{align*}
Therefore, the post-selection mean of $\hat\gamma_{j}  $ is not distorted by instrument selection, and the winner's curse does not arise. Following this line of reasoning, we can also lift the winner's curse in two-sample MRs, as long as we can \textit{create} an unbiased estimator of $\gamma_{j}$ that is independent of the instrument selection event within the same exposure dataset. Carrying forward the heuristic of ``creating independence,'' we {propose} a new two-sample MR framework that fully removes the winner's curse. We start by introducing a randomized instrument selection step that deviates from the hard-thresholding rule in \eqref{eq:hard-thresholding-selection}:

\begin{proc}[\normalfont Randomized instrument selection]\label{step:1 instrument selection}
For each SNP $j=1,2,\ldots,p$, generate a pseudo SNP-exposure association effect $Z_j\sim \mathcal{N}(0,\eta^2)$, and select SNP $j$ if
\begin{align*}
\Big| \frac{\hat{\gamma}_j}{ \sigma_{X_j} } + Z_j \Big| >\lambda .
\end{align*}
Also define the set of selected SNPs as
\begin{align*}
\mathcal{S}_{\lambda} = \Big\{j: S_j >0 ,\ j = 1, 2,\ldots,p \Big\}, \quad \text{where } S_j =  \Big| \frac{\hat{\gamma}_j}{ \sigma_{X_j} } + Z_j \Big| -\lambda.
\end{align*}
Here, $\lambda>0$ is a cutoff value, and $\eta$ is a pre-specified constant that reflects the noise level of the pseudo SNPs. Discussions on the choice of $\lambda$ and $\eta$ are given in Section \ref{Section:simulation-tuning}. 
\end{proc}
It is worth mentioning that although we randomize $\hat{\gamma}_j$ before checking whether it passes the threshold $\lambda$, such randomization will not affect the selection of strong SNPs (i.e., when $|\gamma_j/\sigma_{X_j}| \gg 0$). In other words, the pseudo effects we introduce merely randomize the selection around the cutoff value $\lambda$. The benefit of introducing randomized instrument selection will be apparent in the following bias removal step.
	
\begin{proc}[\normalfont Winner's curse removal by Rao-Blackwellization]\label{step:2 bias correction}
For each selected SNP $j\in \mathcal{S}_\lambda$, construct an unbiased estimator of $\gamma_j$ as
\begin{align*}
\hat{\gamma}_{j,\mathtt{RB}} &= \hat{\gamma}_j - \frac{\sigma_{X_j}}{\eta}\frac{\phi\big(A_{j,+}\big) - \phi\big(A_{j,-}\big)}{1 - \Phi\big(A_{j,+}\big) + \Phi\big(A_{j,-}\big)},\text{ where } A_{j,\pm} =  -  \frac{\hat{\gamma}_j}{ \sigma_{X_j}\eta } \pm\frac{\lambda}{\eta},
\end{align*}
 and $\phi(\cdot)$ and $\Phi(\cdot)$ denote the standard normal density and cumulative distribution functions.
\end{proc}
	
To understand the above construction, consider the crude ``initial estimator,'' $\hat{\gamma}_{j,\mathtt{ini}} = \hat{\gamma}_j - \frac{\sigma_{X_j}}{\eta^2}Z_j$. By standard calculations, it is unbiased for $\gamma_j$, and is also independent of the IV selection event $j\in\mathcal{S}_\lambda$. As a result, $\hat{\gamma}_{j,\mathtt{ini}}$ is unbiased for $\gamma_j$ both before and after IV selection: 
\begin{align*}
\Expectation\left[\left. \hat{\gamma}_{j,\mathtt{ini}} \right|  j\in\mathcal{S}_\lambda \right]  = \Expectation\left[ \hat{\gamma}_{j,\mathtt{ini}} \right]  = \gamma_j.
\end{align*}
Then we employ Rao-Blackwellization to improve the initial estimator (i.e., to reduce its variance) by projecting it onto the sufficient statistic of the selection likelihood: 
\begin{align*}
\hat{\gamma}_{j,\mathtt{RB}} = \Expectation\left[\left. \hat{\gamma}_{j,\mathtt{ini}} \right| \hat{\gamma}_j,\ j\in\mathcal{S}_\lambda \right]. 
\end{align*}
Implied by the Rao-Blackwell theorem, our $\hat{\gamma}_{j,\mathtt{RB}}$ not only is unbiased for the true $\gamma_j$ conditioning on the selection event, but also achieves the minimum variance for a given $\eta$. A rigorous justification of our proposal will be given in Section \ref{Sec:theory}. Before proceeding, we note that the ``initial estimator'' is introduced only for theoretical discussions and to motivate our construction. Our approach neither employs this initial estimator nor requires constructing it explicitly. In a two-stage genome-wide association studies setting, a similar strategy based on Rao-Blackwellization has been adopted in \cite{bowden2009unbiased}.
	
\begin{figure}[!ht]
\centering
\includegraphics[width=\linewidth]{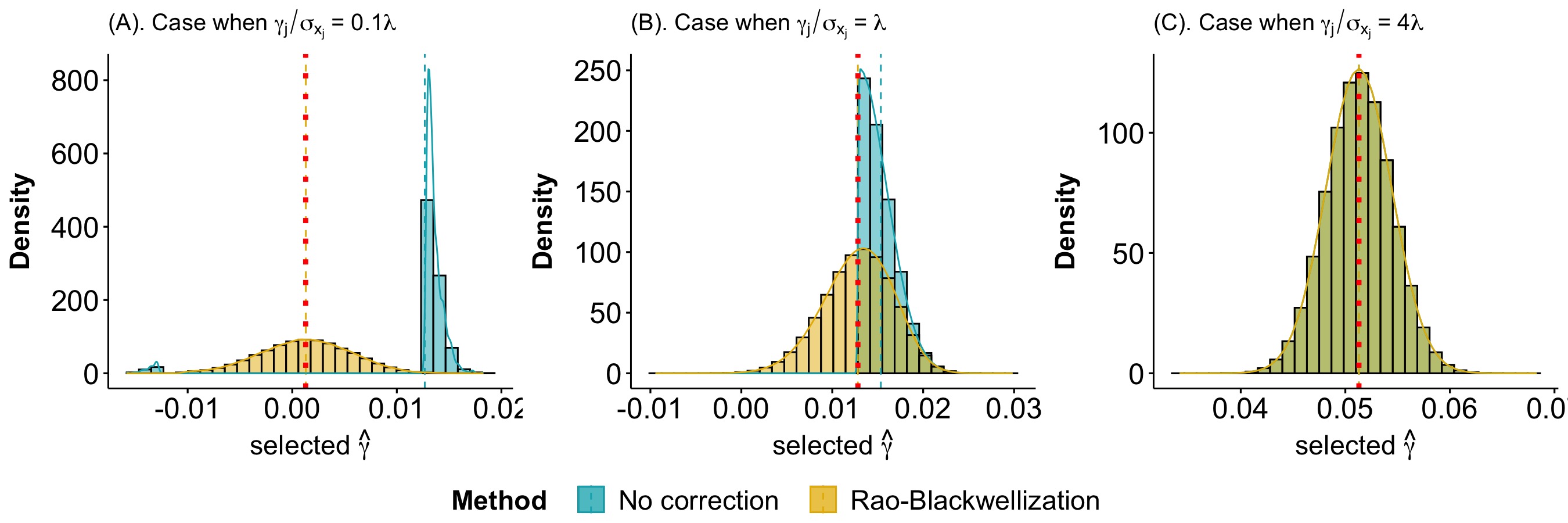}
\caption{Comparison between $\hat{\gamma}_{j,\mathtt{RB}}$ and $\hat{\gamma}_j$ after SNP selection. Panel (A): the weak IV scenario with $\gamma_j/\sigma_{X_j} = 0.1\lambda$. Panel (B): the moderately strong IV scenario with $\gamma_j/\sigma_{X_j} = \lambda$. Panel (C): the strong IV scenario with $\gamma_j/\sigma_{X_j} = 4\lambda$. \xwm{The true instrument effect, $\gamma_j$, is represented by red dotted lines.}}
\label{fig:rb-illustration}
\end{figure}

To illustrate how instrument selection affects the distributions of $\hat{\gamma}_{j}$, and to demonstrate the performance of our bias-corrected $\hat{\gamma}_{j,\mathtt{RB}}$, \xwm{we provide a simulation study in which we generate $\hat{\gamma}_j/\sigma_{X_j}$ from the normal distribution $\mathcal{N}(\gamma_j/\sigma_{X_j}, 1)$. We vary the true normalized instrument effect, $\gamma_j/\sigma_{X_j}$, in the set $\{0.1\lambda,\ \lambda,\ 4\lambda  \}$, representing weak, moderately strong, and strong instruments, respectively.} We set $\lambda = \Phi^{-1}(1- 5\times 10^{-5}/2)$. From the three histograms in Figure \ref{fig:rb-illustration}, we observe that the na\"{i}ve estimator is generally biased since it does not account for selection (unless the given SNP is a very strong instrument). On the other hand, our Rao-Blackwellized estimator remains correctly centered regardless of the strength of the instrument. 

%--------------------------------------------------------------------------------
% Rerandomized IVW estimator
%--------------------------------------------------------------------------------
\subsection{Rerandomized IVW estimator}

Given the unbiased estimate of $\gamma_j$ in the previous section, we propose our final estimator using $\hat{\gamma}_{j,\mathtt{RB}}$. Because Mendelian randomization is the term applied to the random assortment of alleles at the time of gamete formation, and our approach further randomizes instrument/SNP selection, we name $\hat{\beta}_{\mathtt{RIVW}} $ in the following the rerandomized IVW (RIVW) estimator. 
\begin{proc}[\normalfont Rerandomized IVW estimator]\label{step:3 RIVW}
Construct the RIVW estimator via
\begin{align}\label{eq:RIVW}
\hat{\beta}_{\mathtt{RIVW}} &= \frac{\sum_{j\in \mathcal{S}_{\lambda}}  \hat\Gamma_j \hat\gamma_{j,\mathtt{RB}} / {\sigma}_{Y_j}^{2} }
{\sum_{j\in \mathcal{S}_{\lambda}}  (\hat\gamma_{j,\mathtt{RB}}^2 - \hat{\sigma}_{X_j,\mathtt{RB}}^{ \mathrm{2} }  ) / {\sigma}_{Y_j}^{2}}, 
\end{align}
where  
\begin{align*}
\hat{\sigma}_{X_j,\mathtt{RB}}^{ \mathrm{2} }  =&\ \sigma_{X_j}^2\Bigg( 1 - \frac{1}{\eta^2}  \frac{A_{j,+}\phi(A_{j,+}) - A_{j,-}\phi(A_{j,-})}{1 - \Phi(A_{j,+}) + \Phi( A_{j,-} )} +\frac{1}{\eta^2} \Big(\frac{\phi(A_{j,+}) - \phi(A_{j,-}) }{1 - \Phi(A_{j,+}) + \Phi( A_{j, -} )}\Big)^2 \Bigg),
\end{align*}
and $A_{j,\pm}$ are defined in Step \ref{step:2 bias correction}.
\end{proc}

The specific form of our estimator deviates from the classical IVW estimator defined in \eqref{eq:IVW-classical}. Such deviation is motivated by the consideration that $\hat\gamma_{j,\mathtt{RB}}$ are random quantities measured with errors. We subtract from $\hat\gamma_{j,\mathtt{RB}}^2$ a variance estimate, $\hat{\sigma}_{X_j,\mathtt{RB}}^2$, to correct this measurement error bias. 

We are now ready to present our final step, statistical inference in two-sample MR with summary data accounting for both instrument selection bias and measurement error bias. 

\begin{proc}[\normalfont Statistical inference]\label{step:4 inference}
Let the estimated variance be
\begin{align}\label{eq:variance-estimator}
\hat{\mathsf{V}}_{\mathtt{RIVW}} &= \frac{\sum_{j\in S_{\lambda}}  \left(\hat\Gamma_j \hat\gamma_{j,\mathtt{RB}} -  \hat{\beta}_{\mathtt{RIVW}}  (\hat\gamma_{j,\mathtt{RB}}^{\mathrm{2}} - \hat{\sigma}_{X_j,\mathtt{RB}}^{\mathrm{2}} ) \right)^2/\sigma_{Y_j}^4}{\left(\sum_{j\in S_{\lambda}}  \Big(\hat\gamma_{j, \mathtt{RB}}^{\mathrm{2}} - \hat{\sigma}_{X_j,\mathtt{RB}}^{\mathrm{2} }\Big)/\sigma_{Y_j}^2\right)^2},
\end{align}
then a level $1-\alpha$ confidence interval can be constructed as
\begin{align*}
\left[\ \hat{\beta}_{\mathtt{RIVW}} - \Phi^{-1}(1-\frac{\alpha}{2}) \sqrt{\hat{\mathsf{V}}_{\mathtt{RIVW}}}\ ,\ \ \hat{\beta}_{\mathtt{RIVW}} + \Phi^{-1}(1-\frac{\alpha}{2}) \sqrt{\hat{\mathsf{V}}_{\mathtt{RIVW}}}\ \right].
\end{align*}
\end{proc}

Because our RIVW estimator resembles the ``slope coefficient''  obtained from a regression of $\hat{\Gamma}_j$ on $\hat{\gamma}_{j,\mathtt{RB}}$, the proposed variance estimator can be motivated from the use of ``regression residuals'' (See Section \ref{Section:measurement-error-bias} for detailed discussion). Thanks to the regression interpretation of the RIVW estimator, and because the presence of balanced horizontal pleiotropy simply adds centered noise to the SNP-outcome association estimates $\hat{\Gamma}_j$, we shall demonstrate in Section \ref{Sec:balanced horizontal pleiotropy} that our RIVW estimator remains asymptotically unbiased and normally distributed under the more general Assumption \ref{assump:Measurement-error-pleiotropy}$^\prime$, and our variance estimator is valid without further modification. 

In Section \ref{section:smoothing-over-randomization}, we consider an extension to our RIVW estimator, where we further ``smooth'' over the random IV selection, leading to an estimator that re-weights the SNP-exposure association estimates. Finally, Section \ref{Sec:extension-correlated-IV} provides a first attempt to correct LD clumping bias by generalizing our randomized instrument selection and Rao-Blackwellization to a correlated instrumental variables setting. 

%%%%%%%%%%%%%%%%%%%%%%%%%%%%%%%%%%%%%%%%%%%%%%%%%%%%%%%%%%%%%%%%%%%%%%%%%%%%%%%%%
% Theoretical investigations
%%%%%%%%%%%%%%%%%%%%%%%%%%%%%%%%%%%%%%%%%%%%%%%%%%%%%%%%%%%%%%%%%%%%%%%%%%%%%%%%% 
\section{Theoretical investigations}\label{Sec:theory}
	
In this section, we first provide theoretical justifications for our winner's curse removal procedure. We then demonstrate the statistical validity of our inference approach by showing that the RIVW estimator is asymptotically normally distributed under mild conditions (Theorem \ref{thm:asy normality}). Finally, we show that our variance estimator is consistent for the leading variance of the RIVW estimator (Theorem \ref{thm:valid standard error}).

%--------------------------------------------------------------------------------
% Notation and assumptions
%--------------------------------------------------------------------------------	
\subsection{Notation and assumptions}
	
Before discussing the theoretical results in detail, we revisit and introduce some notation and assumptions adopted in the paper. Recall that the $j$th genetic instrument/SNP is selected if and only if $S_j = |\frac{\hat{\gamma}_j}{\sigma_{X_j}} + Z_j| - \lambda > 0$.  The collection of selected instruments is denoted by $\mathcal{S}_\lambda = \{j:S_j > 0\}$. The  asymptotic regime we consider involves $p\to\infty$ and $\lambda\to \infty$. This captures the phenomena of ``many instruments/SNPs'' in MR studies. On the other hand, the requirement that $n_X,n_Y\to \infty$ is implicit, as we follow the literature and assume that the estimated SNP-exposure and SNP-outcome associations are normally distributed (Assumption \ref{assump:Measurement-error}). We also introduce the following notation for probabilistic ordering. For two (sequences) of random variables, $A$ and $B$, write $A \precsimProb B$ if the ratio $A/B$ is asymptotically bounded in probability. The strict relation, $A\precProb B$, implies that $A/B\toProb 0$. Finally, $A\asympProb B$ indicates both $A\precsimProb B$ and $B\precsimProb A$. 
	
To facilitate discussion, we introduce two additional quantities: the number of selected instruments as $p_\lambda = |\mathcal{S}_\lambda |$, and an overall measure of instrument strength after selection:
\begin{align*}
\kappa_\lambda &= \frac{1}{p_\lambda}\sum_{j \in \mathcal{S}_\lambda}\left( \frac{\gamma_j}{\sigma_{X_j}} \right)^2.
\end{align*}
We note that both $p_\lambda$ and $\kappa_\lambda$ are random, as they depend on the set of selected instruments. In addition, $\kappa_\lambda$ is typically unknown, because it depends on the unobserved SNP-exposure associations, $\gamma_j$. 

In addition to Assumption \ref{assump:Measurement-error}, we employ the following conditions.  
	
\begin{assumption}[\normalfont Instrument selection]\label{assump:instrument selection}  The cutoff value satisfies $\lambda \to \infty$.
\end{assumption}

\begin{assumption}[\normalfont No dominant instrument]\label{assump: No dominating instrument}
The true instrument effect satisfies 
$$\max_{j\in S_{\lambda}} \gamma_j^2\ \Big/\ \Big(\sum_{j\in S_{\lambda}} \gamma_j^2\Big)\toProb 0.$$
\end{assumption}
    
Assumption \ref{assump:instrument selection} requires the cutoff value to diverge, which is quite plausible given that $\lambda$ is usually of the order $\sqrt{\log p}$ to account for multiple testing; see footnote \ref{footnote:1}. Assumption \ref{assump: No dominating instrument} requires that, after selection, no instrument has a ``dominating effect.'' It helps rule out the extreme scenario where only a handful of genetic variants are relevant. We will employ this condition when establishing the asymptotic normality of our estimator.\footnote{To better explain what Assumption \ref{assump: No dominating instrument} entails, consider the simulation setting in Section \ref{Section:simulation-no-pleiotropy}, and ignore the instrument selection step for simplicity. Then it is easy to show that $\max_{1\leq j\leq p}\gamma_j^2 \precsim \varepsilon_x^2\log(p\pi_x)$, while on the other hand $\sum_{j=1}^p \gamma_j^2 \asymp \varepsilon_x^2 p\pi_x$.}  

%--------------------------------------------------------------------------------
% Properties of the Rao-Blackwellized estimator
%--------------------------------------------------------------------------------	
\subsection{Properties of the Rao-Blackwellized estimator}\label{Sec:RB-theorey}
	
Our first result verifies that the proposed framework breaks the winner's curse: the construction, $\hat{\gamma}_{j,\mathtt{RB}}$, is unbiased for $\gamma_j$ conditional on the selection event $j\in \mathcal{S}_\lambda$. 
	
\begin{lem}[\normalfont Instrument selection bias correction]\label{lem:Rao-Blackwellization}
Under Assumption \ref{assump:Measurement-error},  $\Expectation[\hat{\gamma}_{j,\mathtt{RB}}|j\in\mathcal{S}_\lambda]=\gamma_j$.
\end{lem}
	
As an immediate result of this lemma, $\hat{\gamma}_{j,\mathtt{RB}}$ is preferred to $\hat{\gamma}_j$ since it helps lift the winner's curse by correcting instrument selection bias. To provide some intuition, the unbiasedness of $\hat{\gamma}_{j,\mathtt{RB}}$ stems from properties of the initial estimator, $\hat{\gamma}_{j,\mathtt{ini}}$, which we introduced in the previous section. In particular, the initial estimator is unbiased due to its independence from the selection event. Therefore, because $\hat{\gamma}_{j,\mathtt{RB}}=\Expectation[\hat{\gamma}_{j,\mathtt{ini}}|\hat{\gamma}_j,j\in\mathcal{S}_\lambda]$, iterated expectation implies $\Expectation[\hat{\gamma}_{j,\mathtt{RB}}|j\in\mathcal{S}_\lambda] = \Expectation[\hat{\gamma}_{j,\mathtt{ini}}|j\in\mathcal{S}_\lambda] = \gamma_j$. Finally, the specific expression of $\hat{\gamma}_{j,\mathtt{RB}}$ can be found by explicit calculation, which is available in the Supplementary Material \citep{Ma-Wang-Wu-2022-AOS-Supp}. On a related note, $\hat{\gamma}_{j,\mathtt{RB}}$ also has an advantage over the initial estimator: although the initial estimator is unbiased, $\hat{\gamma}_{j,\mathtt{RB}}$ has a smaller variance thanks to Rao-Blackwellization. 

As we will discuss in the next subsection, correcting measurement error bias requires the variance of the estimated SNP-exposure associations. In conventional MR analyses without instrument selection or when the selection step is carried out in an independent third sample, this information is directly available from $\sigma_{X_j}^2$. In our framework, however, both the selection step and the subsequent Rao-Blackwellization may affect the variance of the selected instrument effects. That is, $\Var[\hat{\gamma}_{j,\mathtt{RB}}|j\in\mathcal{S}_\lambda]\neq \sigma_{X_j}^2$ in general. We characterize this conditional variance in the following lemma. 

\begin{lem}[\normalfont Variance of $\hat{\gamma}_{j,\mathtt{RB}}$]\label{lem:variance of gamma_RB}
Under Assumption \ref{assump:Measurement-error}, conditioning on the selection event, the variance of $\hat{\gamma}_{j,\mathtt{RB}}$ is
\xwm{\begin{align*}
\sigma_{X_j,\mathtt{RB}}^{ \mathrm{2} } &= \sigma_{X_j}^2\Bigg( 1 - \frac{1}{\eta\Prob[S_j > 0]}\int_{-\infty}^\infty y\phi(y)\Big(\phi\big(B_{j,+}(y)\big) - \phi\big(B_{j,-}(y)\big)\Big)  \diff y \\
&\qquad \qquad \qquad + \frac{1}{\eta^2\Prob[S_j > 0]}\int_{-\infty}^\infty \phi(y)\frac{\Big(\phi\big(B_{j,+}(y)\big) - \phi\big(B_{j,-}(y)\big)\Big)^2}{1-\Phi\big(B_{j,+}(y)\big) + \Phi\big(B_{j,-}(y)\big)}  \diff y\Bigg), 
\end{align*}
where we define $B_{j,\pm}(y) = -\left(\frac{{\gamma}_j}{\sigma_{X_j}\eta} + \frac{y}{\eta}\right) \pm \frac{\lambda}{\eta}$. }
\end{lem}

The above expression clearly demonstrates that the conditional variance depends on the true SNP-exposure associations, $\gamma_j$, in a nonlinear and complicated way. As a result, a direct plug-in estimator of the conditional variance is not expected to perform well. 

To provide some intuition for our variance estimator in Step \ref{step:3 RIVW}, we recall that the Rao-Blackwellized instrument effects are obtained as conditional expectations. Therefore, instead of characterizing its variance directly, one can employ the conditional variance decomposition formula, that is, 
\begin{align*}
\sigma_{X_j,\mathtt{RB}}^2 = \Var\left[\left. \hat{\gamma}_{j,\mathtt{ini}} \right| j\in\mathcal{S}_\lambda \right] - \Expectation\left[\left.\Var\left[\left. \hat{\gamma}_{j,\mathtt{ini}} \right| \hat{\gamma}_j,j\in\mathcal{S}_\lambda \right]\right| j\in\mathcal{S}_\lambda \right].
\end{align*}
Due to independence, the first term on the right side is simply the unconditional variance of the initial estimator: $\Var\left[\left. \hat{\gamma}_{j,\mathtt{ini}} \right| j\in\mathcal{S}_\lambda \right] = \Var[\hat{\gamma}_{j,\mathtt{ini}} ]$. The second term, however, is much more difficult to obtain. In fact, if one computes the outer expectation, then it reduces to the formula in Lemma \ref{lem:variance of gamma_RB}. Our insight is that we define the variance estimator without explicitly computing the outer expectation. To be more precise, $\hat{\sigma}_{X_j,\mathtt{RB}}^2$ is defined from 
\begin{align*}
\hat{\sigma}_{X_j,\mathtt{RB}}^2 &= \Var[\hat{\gamma}_{j,\mathtt{ini}} ] - \Var\left[\left. \hat{\gamma}_{j,\mathtt{ini}} \right| \hat{\gamma}_j,j\in\mathcal{S}_\lambda \right].
\end{align*}
Due to iterative expectation, the above is unbiased for $\sigma_{X_j,\mathtt{RB}}^2$ (conditional on the selection event), but is generally inconsistent. Fortunately, for measurement error bias correction, we only need to estimate the aggregate variance, $\sum_{j\in\mathcal{S}_\lambda}\sigma_{X_j,\mathtt{RB}}^2$. We further discuss this issue below. 

%--------------------------------------------------------------------------------
% Measurement error bias correction
%--------------------------------------------------------------------------------
\subsection{Measurement error bias correction}\label{Section:measurement-error-bias}

The preceding section illustrates how our approach breaks the winner's curse. We now turn to discussing how our RIVW estimator also removes measurement error bias, which arises due to the randomness in $\hat{\gamma}_j$ (or $\hat{\gamma}_{j,\mathtt{RB}}$). The measurement error bias issue can be heuristically understood by viewing the IVW estimator as an estimated regression coefficient. In a hypothetical case where $\hat{\gamma}_j$ are measured without any error (i.e., $\hat{\gamma}_j = \gamma_j$, also referred to as no measurement error assumption in the literature, \cite{bowden2016assessing}), an ``oracle'' IVW estimator would regress $\hat{\Gamma}_j$ on ${\gamma}_j$ (after attaching a weight $1/\sigma_{Y_j}$ to them) and obtain the slope coefficient as an estimator for $\beta$. Such an oracle IVW estimator is unbiased for $\beta$. In practice, however, we only observe $\gamma_j$ with a random error, which means that the slope estimate from regressing $\hat{\Gamma}_j$ on $\hat{\gamma}_j$ will be biased.  Such a bias is also referred to as regression attenuation, or weak instrument bias in the MR literature \citep{sadreev2021navigating}.

Our RIVW removes measurement error bias following a similar logic to the strategy adopted in \cite{ye2019debiased}, who propose to replace $\hat{\gamma}_j^2$ with $\hat{\gamma}_j^2 - \sigma_{X_j}^2$ in the denominator of the conventional IVW estimator. Despite taking a similar form, we propose the use of a different variance estimator, $\hat{\sigma}_{X_j,\mathtt{RB}}^2$ (Step \ref{step:3 RIVW}), because the Rao-Blackwellization step can affect the variance of the instrument effects. That is, $\Var[\hat{\gamma}_{j,\mathtt{RB}}|j\in\mathcal{S}_\lambda]\neq   \sigma_{X_j}^2 = \Var[\hat{\gamma}_j]$. Although Lemma \ref{lem:variance of gamma_RB} provides a formula for the variance of the Rao-Blackwellized estimator, $\hat{\gamma}_{j,\mathtt{RB}}$, the result is not directly amenable to implementation as the variance formula depends on the unknown $\gamma_j$ nonlinearly. Nevertheless, thanks to the special form of the RIVW estimator, we only need to correctly estimate the ``aggregated variance,'' $\sum_{j \in S_\lambda } \sigma_{X_j,\mathtt{RB}}^{ \mathrm{2} }$, to remove measurement error bias, as the next lemma demonstrates.

\begin{lem}[\normalfont Variance estimation]\label{lem:variance estimation}
Assume Assumptions \ref{assump:Measurement-error} and \ref{assump:instrument selection} hold. Then $\Expectation[\hat{\sigma}_{X_j,\mathtt{RB}}^2|\mathcal{S}_{\lambda}]=\sigma_{X_j,\mathtt{RB}}^2$, and
\begin{align*}
\left|\sum_{j \in S_\lambda } \hat{\sigma}_{X_j,\mathtt{RB}}^2 - \sum_{j \in S_\lambda } \sigma_{X_j,\mathtt{RB}}^2\right| \precsimProb \sqrt{p_\lambda}\nu^2\lambda.
\end{align*}
\end{lem}

We will postpone the discussion on the above probabilistic order to the next subsection after we introduce the main asymptotic normality result.

%--------------------------------------------------------------------------------
% Properties of the RIVW estimator
%--------------------------------------------------------------------------------
\subsection{Properties of the RIVW estimator}

We are now in a position to describe the asymptotic behavior of our RIVW estimator.

\begin{thm}{\normalfont{(Asymptotic normality)}}\label{thm:asy normality}
Assume Assumptions \ref{assump:Measurement-error}, \ref{assump:instrument selection}, and \ref{assump: No dominating instrument} hold, $p_\lambda\toProb \infty$, and $\kappa_\lambda/\lambda^2\toProb \infty$. Then conditional the selection event $\mathcal{S}_{\lambda}$,
\begin{align*}
\mathsf{V}_{\mathtt{RIVW}}^{-\frac{1}{2}}\left(\hat{\beta}_{\mathtt{RIVW}} - \beta\right) \overset{\mathcal{D}}{\to} \mathcal{N}(0,1),
\end{align*}
where 
\begin{align*}
\mathsf{V}_{\mathtt{RIVW}} &= \frac{\Var\left[ \sum_{j\in S_{\lambda}}u_{j,\mathtt{RIVW}}/\sigma_{Y_j}^2 \Big| S_\lambda \right]}{\left(\sum_{j\in S_{\lambda}}\gamma_j^2 /\sigma_{Y_j}^2\right)^2} \asympProb \frac{1}{p_\lambda \kappa_\lambda},\\
u_{j,\mathtt{RIVW}} &= \gamma_j\Big(u_{Y_j} - \beta u_{X_j,\mathtt{RB}}\Big) +   \Big(u_{X_j,\mathtt{RB}}u_{Y_j} - \beta(u_{X_j,\mathtt{RB}}^{2} - \hat{\sigma}_{X_j,\mathtt{RB}}^{2})\Big),
\end{align*}
$u_{Y_j} = \hat{\Gamma}_j - \Gamma_j$, and $u_{X_j,\mathtt{RB}} = \hat{\gamma}_{j,\mathtt{RB}} - \gamma_j$.
\end{thm}

We provide a decomposition below to demonstrate the intuition behind the asymptotic unbiasedness of our RIVW estimator and defer the formal proof to the Supplementary Material \citep{Ma-Wang-Wu-2022-AOS-Supp}. This decomposition also sheds light on the mechanism of our measurement error bias correction and how the error from variance estimation  features in the asymptotic analysis. 

To start, we rewrite our RIVW estimator as
\begin{align*}
\hat{\beta}_{\mathtt{RIVW}} 
&= \beta 
+ \frac{\sum_{j\in \mathcal{S}_{\lambda}}  u_{j,\mathtt{RIVW}} /\sigma_{Y_j}^2}{\sum_{j\in \mathcal{S}_{\lambda}}  (\hat\gamma_{j,\mathtt{RB}}^{2} - \hat{\sigma}_{X_j,\mathtt{RB}}^{2} )/\sigma_{Y_j}^2}.
\end{align*}
Then from Lemmas \ref{lem:Rao-Blackwellization} and \ref{lem:variance of gamma_RB}, it should be clear that $u_{j,\mathtt{RIVW}}$ has a zero mean, which implies that our RIVW estimator is asymptotically unbiased. Lemma \ref{lem:variance estimation} further helps demonstrate that the denominator in our RIVW estimator converges to $\sum_{j\in \mathcal{S}_{\lambda}}  \gamma_{j}^{2} /\sigma_{Y_j}^2$. 

It is not surprising that the asymptotic variance of our RIVW estimator takes a complicated form, as the variance of $u_{j,\mathtt{RIVW}}$ involves higher order moments of the Rao-Blackwellized instrument effects, $\hat{\gamma}_{j,\mathtt{RB}}$. We employ the condition $\kappa_\lambda/\lambda^2\toProb \infty$ to simplify the analysis. In particular, this assumption implies that the leading variance stems from the first term in the expression of $u_{j,\mathtt{RIVW}}$. This condition seems mild and quite plausible in applications since instrument selection will help improve the overall IV strength. See the Supplemental Material for additional discussions and results on the probabilistic order of $\kappa_\lambda$. 

The previous decomposition also motivates our regression residual based variance estimator. In particular, we note that the ``error term,'' $u_{j,\mathtt{RIVW}}$, can also be written as $\hat\Gamma_j \hat\gamma_{j,\mathtt{RB}} -  \beta  (\hat\gamma_{j,\mathtt{RB}}^2 - \hat{\sigma}_{X_j,\mathtt{RB}}^2 )$. As a result, in our standard error construction, we replace the unknown terms by their estimates. The following theorem establishes the consistency of our standard error.

\begin{thm}{\normalfont{(Consistent variance estimation)}}\label{thm:valid standard error}
Assume Assumptions \ref{assump:Measurement-error}, \ref{assump:instrument selection}, and \ref{assump: No dominating instrument} hold, $p_\lambda\toProb \infty$, and $\kappa_\lambda/\lambda^2\toProb \infty$. Then 
\begin{align*}
\frac{\hat{\mathsf{V}}_{\mathtt{RIVW}}}{{\mathsf{V}}_{\mathtt{RIVW}}}\toProb 1.
\end{align*}
\end{thm}

%%%%%%%%%%%%%%%%%%%%%%%%%%%%%%%%%%%%%%%%%%%%%%%%%%%%%%%%%%%%%%%%%%%%%%%%%%%%%%%%%
% Extensions
%%%%%%%%%%%%%%%%%%%%%%%%%%%%%%%%%%%%%%%%%%%%%%%%%%%%%%%%%%%%%%%%%%%%%%%%%%%%%%%%% 
\section{Extensions}\label{Sec:Extensions}

%--------------------------------------------------------------------------------
% Balanced horizontal pleiotropy
%--------------------------------------------------------------------------------
\subsection{Balanced horizontal pleiotropy}\label{Sec:balanced horizontal pleiotropy}
		
Balanced pleiotropy (also known as systematic pleiotropy) refers to the scenario that the exclusion restriction assumption breaks down and the genetic variants have equal chances to affect the outcome either positively or negatively \citep{hemani2018evaluating,xue2021constrained}. For this extension, we consider the following model that is commonly adopted in the literature:
\xwm{
\begin{align}
&Y = \beta X +\sum_{j=1}^p \alpha_j G_j + \theta_Y U + E_Y, \quad \text{for each}\ j = 1, \ldots, p.
\end{align}
}
In the above, $\alpha_j$ captures the pleiotropic effect of SNP $j$ on the outcome $Y$. Formally, we require
\begin{assumptionalt}[\normalfont Measurement error model under balanced horizontal pleiotropy]\label{assump:Measurement-error-pleiotropy}
(i) For any $j\neq j'$, the pairs $(\hat\Gamma_j,\hat\gamma_j)$ and $(\hat\Gamma_{j'},\hat\gamma_{j'})$ are mutually independent.\\
(ii) For each $j$, 
\begin{align*}
\begin{bmatrix}
\hat{\Gamma}_j\\
\hat\gamma_j
\end{bmatrix} \sim \mathcal{N}\left(\begin{bmatrix}
\beta\gamma_j + \alpha_j \\
\gamma_j
\end{bmatrix}\ ,\ \begin{bmatrix}
{\sigma}_{Y_j}^2 & 0 \\
0 & {\sigma}_{X_j}^2
\end{bmatrix}\right),
\end{align*}
In addition, there exists some $\nu\to 0$, such that $\{\sigma_{Y_j}/\nu,\sigma_{X_j}/\nu:1\leq j\leq p\}$ are uniformly bounded and bounded away from zero. \\
\xwm{(iii) The pleiotropic effects, $\alpha_1, \ldots, \alpha_p$, are mutually independent and follow a distribution with mean 0, variance $\tau^2$, and bounded third moment. In addition, $\tau/\nu$ is bounded.}
\end{assumptionalt}
	
Because the pleiotropic effects follow a centered distribution, they do not introduce any additional bias into our RIVW estimator. In fact, as the reader will gather from checking our proof, the presence of balanced pleiotropic effects merely inflates the variance of the RIVW estimator. That is, our RIVW estimator remains asymptotically unbiased and normal under the more general Assumption \ref{assump:Measurement-error-pleiotropy}$^\prime$. Thanks to our regression residual based variance estimator in \eqref{eq:variance-estimator}, the statistical inference procedure in Step \ref{step:4 inference} remains valid in this setting. In other words, our variance estimator can be directly applied to the balanced horizontal pleiotropy scenario without any modification. This feature allows practitioners to conduct statistical analysis without switching among various variance estimators under different assumptions on the existence of balanced horizontal pleiotropy. We also note that this balanced horizontal pleiotropy assumption can be viewed as a sufficient condition of Instrument Strength Independent on Direct Effect (InSIDE) assumption, requiring that the exposure effects of individual SNPs are independent of their pleiotropic effects on the outcome \citep{burgess2017interpreting}.
	
We summarize this conclusion in the following corollary: 
\begin{coro}{\normalfont{(Validity of the RIVW estimator under balanced horizontal pleiotropy)}}\label{coro:pleiotropy}
The results in Theorems \ref{thm:asy normality} and \ref{thm:valid standard error} continue to hold if Assumption \ref{assump:Measurement-error} is replaced by \ref{assump:Measurement-error-pleiotropy}$^\prime$. 
\end{coro}
	
%--------------------------------------------------------------------------------
% Smoothing over randomization
%--------------------------------------------------------------------------------
\subsection{Smoothing over randomization}\label{section:smoothing-over-randomization} 
	
Our RIVW estimator is constructed based on the selected instruments. While the selection step might be of interest, as it allows the researcher to further evaluate the validity of the selected instruments and the underlying causal mechanism, we do recognize that introducing the pseudo SNP-exposure effects, $Z_j$, into the procedure may inflate the variance of the RIVW estimator. In this subsection, we consider an extension where each instrument is weighted by its (conditional) probability of being selected. Formally, we define
\begin{align*}
\hat{\omega}_j &= \Prob\left[ j\in\mathcal{S}_\lambda | \hat{\gamma}_j \right] = 1 - \Phi\left(A_{j,+}\right) + \Phi\left(A_{j,-}\right),
\end{align*}
and we consider the following smoothed RIVW (sRIVW) estimator
\begin{align*}
\hat{\beta}_{\mathtt{sRIVW}} 
&= \frac{\sum_{j = 1}^p \hat{\Gamma}_j\hat{\gamma}_{j,\mathtt{RB}}\hat{\omega}_j/\sigma_{Y_j}^2}{\sum_{j = 1}^p (\hat{\gamma}_{j,\mathtt{RB}}^2 - \hat{\sigma}_{X_j,\mathtt{RB}}^2)\hat{\omega}_j/\sigma_{Y_j}^2} .
\end{align*}
To conserve space, formal results, such as asymptotic normality and consistent variance estimation, are collected in the Supplementary Material \citep{Ma-Wang-Wu-2022-AOS-Supp}. In what follows, we intuitively demonstrate the properties of the sRIVW estimator. 

To start, we consider the expectations of the numerator and denominator of the sRIVW estimator. By iterative expectation, it is straightforward to show that
\begin{align*}
\Expectation\left[\hat{\gamma}_{j,\mathtt{RB}}\hat{\omega}_j\right] &= \Expectation\left[\hat{\gamma}_{j,\mathtt{RB}}\Indicator_{j\in \mathcal{S}_\lambda}\right] = \gamma_j\omega_j,\quad \omega_j = 1 - \Phi\left(\frac{\lambda - \frac{\gamma_j}{\sigma_{X_j}}}{\sqrt{1 + \eta^2}}\right) + \Phi\left(\frac{-\lambda - \frac{\gamma_j}{\sigma_{X_j}}}{\sqrt{1 + \eta^2}}\right),
\end{align*}
and $\omega_j$ is the unconditional probability of the $j$th instrument being selected. As a result, the numerator of the sRIVW estimator has expectation $\beta\sum_{j=1}^p\gamma_j^2\omega_j/\sigma_{Y_j}^2$. From Lemma \ref{lem:variance estimation} in the previous section, $\hat{\sigma}_{X_j,\mathtt{RB}}^2$ is unbiased for ${\sigma}_{X_j,\mathtt{RB}}^2$ conditional on the selection event, which implies that the denominator of our sRIVW estimator has expectation $\sum_{j=1}^p\gamma_j^2\omega_j/\sigma_{Y_j}^2$. As a result, we expect that our sRIVW approach also delivers accurate estimates of the causal parameter. Evidence on its finite sample performance is provided in Section \ref{Sec:simulation}.  

To close this subsection, we provide in the following an estimator for the variance of the sRIVW estimator. 
 \begin{align*}
\hat{\mathsf{V}}_{\mathtt{sRIVW}} &= \frac{\sum_{j=1}^p   \left(  \hat\Gamma_j \hat\gamma_{j,\mathtt{RB}}\hat{\omega}_j -  \hat{\beta}_{\mathtt{sRIVW}}  (\hat\gamma_{j,\mathtt{RB}}^{\mathrm{2}} - \hat{\sigma}_{X_j,\mathtt{RB}}^{\mathrm{2}} )\hat{\omega}_j \right)^2/\sigma_{Y_j}^4}{\left(\sum_{j=1}^p  \Big(\hat\gamma_{j, \mathtt{RB}}^{\mathrm{2}} - \hat{\sigma}_{X_j,\mathtt{RB}}^{\mathrm{2} }\Big)\hat{\omega}_j/\sigma_{Y_j}^2\right)^2}.
\end{align*}

%--------------------------------------------------------------------------------
% Correlated instrumental variables
%--------------------------------------------------------------------------------
\subsection{Correlated instrumental variables}\label{Sec:extension-correlated-IV}	

With correlated instruments, we consider two remedies to relax the independence assumption adopted in Assumption \ref{assump:Measurement-error}. The first remedy relies on a revised LD pruning procedure (referred to as sigma-based LD pruning), and our proposed RIVW estimator along with its variance estimator remains valid without further modification. As opposed to the classical clumping that sorts SNPs with their p-values, sigma-based LD pruning sorts the correlated genetic variants by their standard deviations, which are often assumed to be known in two-sample MR with summary data. %\citep{ye2019debiased,zhao2020statistical}. 
We then take the first SNP (i.e., the SNP with the smallest standard deviation) and remove all SNPs in linkage disequilibrium with this first SNP.  Our procedure goes on with the next SNP with the smallest standard deviation that has not been removed yet. In the end, the sigma-based LD pruning delivers a collection of independent instruments. In addition, because this revised sigma-based LD pruning procedure does not touch the information in the estimated SNP-exposure effect sizes, no additional selection bias is introduced. We compare the performance of this sigma-based LD pruning with classical LD pruning and p-value based LD clumping in the Supplementary Material \citep{Ma-Wang-Wu-2022-AOS-Supp}.

We propose another remedy relying on a carefully crafted initial estimator that is independent with the classical p-value based LD clumping selection event. We then propose an updated Rao-Blackwellized estimator customized to LD clumping events when IVs are correlated. To conserve space in the main paper, we leave the detailed discussion and implementation of this estimator to the Supplementary Material \citep{Ma-Wang-Wu-2022-AOS-Supp}. 

%%%%%%%%%%%%%%%%%%%%%%%%%%%%%%%%%%%%%%%%%%%%%%%%%%%%%%%%%%%%%%%%%%%%%%%%%%%%%%%%%
% Simulation
%%%%%%%%%%%%%%%%%%%%%%%%%%%%%%%%%%%%%%%%%%%%%%%%%%%%%%%%%%%%%%%%%%%%%%%%%%%%%%%%% 	
\section{Simulation}\label{Sec:simulation}
	
%--------------------------------------------------------------------------------
% Choice of tuning parameters
%--------------------------------------------------------------------------------
\subsection{Choice of tuning parameters}\label{Section:simulation-tuning} 
	
The RIVW estimator depends on the selection threshold ($\lambda$) as well as the level of randomization introduced via the pseudo SNPs ($\eta$). In this section, we discuss practical considerations in choosing these two tuning parameters. To begin with, the choice of the cutoff value $\lambda$ is subjective but needs to be large enough for the selected SNPs to be informative IVs. \xwm{Because the RIVW estimator lifts the winner's curse and removes measurement error bias, our framework allows for including more genetic instruments with moderate effects compared with traditional MR analyses.} Therefore, we recommend $\lambda = \Phi^{-1}(1-\alpha/2)$ with $\alpha = 5\times 10^{-5}$ as the default setting for our RIVW estimator. 
    
\begin{figure}[!htbp]
\centering
\includegraphics[width=\linewidth]{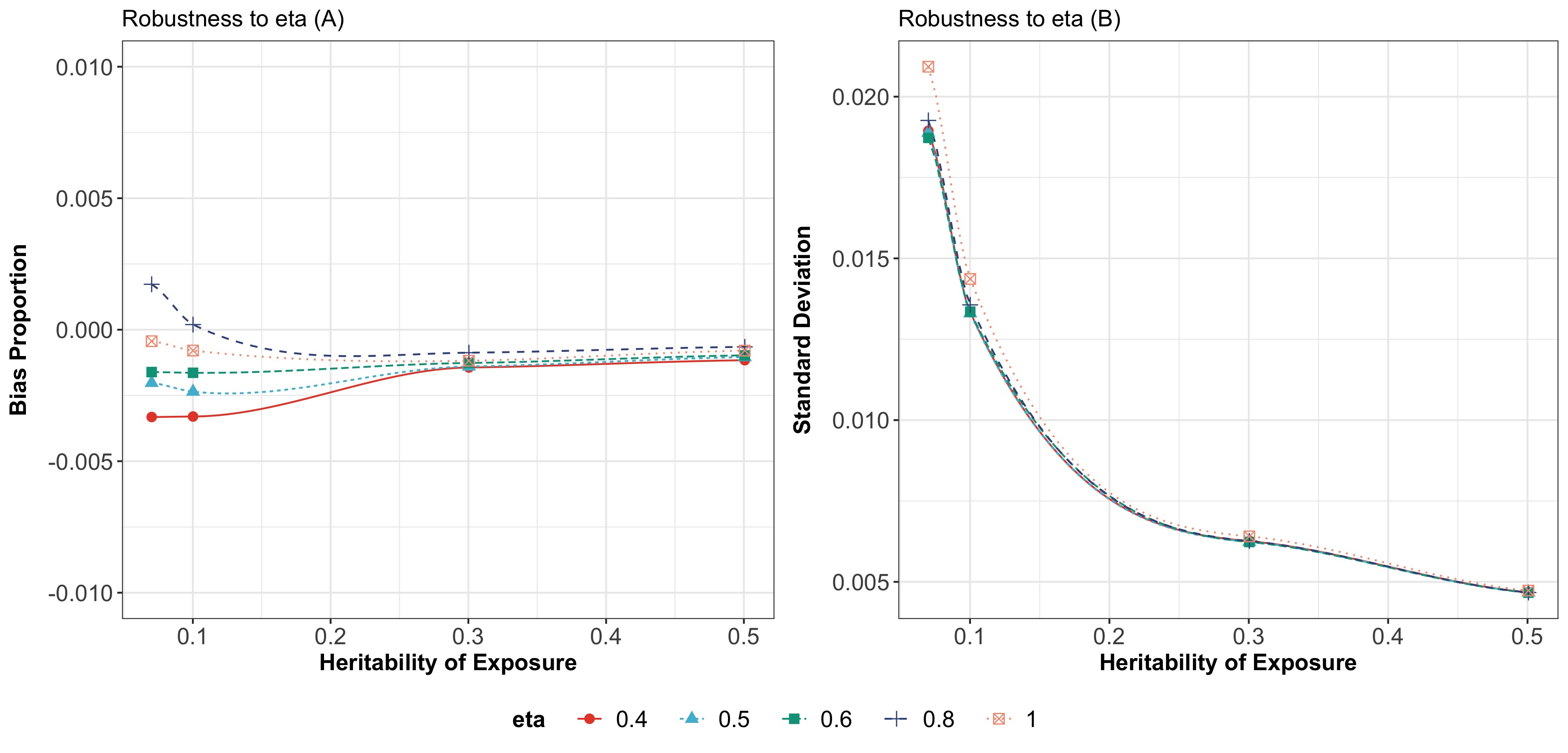}
\caption{\xwm{We generate Monte Carlo samples following the mixture model in Section \ref{Section:simulation-no-pleiotropy} with $\rho=1$ and $\pi_x=\pi_y = 0.005$. We vary $\epsilon_x^2 = \tau^2 $ in the set $\{7\times 10^{-5}, 1\times 10^{-4}, 3\times 10^{-4}, 5\times 10^{-4}\}$. }}
\label{fig:etaselection}
\end{figure}

In terms of choosing $\eta$, our theory only requires it to be bounded away from positive infinity and zero. The reason is quite straightforward: if $\eta$ tends to zero, the pseudo SNPs are asymptotically negligible, which defeats their purposes; on the other hand, if $\eta$ diverges to infinity, the pseudo SNPs mask the original signals $\hat{\gamma}_j/\sigma_{X_j}$ and all instruments will be selected as a result. Rather than studying an optimal choice of $\eta$ via theoretical derivations, extensive simulation evidence suggests that our procedure is not very sensitive to this choice. In Figure \ref{fig:etaselection}, we provide the Monte Carlo bias and standard deviation of our RIVW estimator for different values of $\eta$ in the set $\{0.4,\ 0.5,\ 0.6,\ 0.8,\ 1\}$. We observe that the bias is negligible across different choices of $\eta$. This is in line with our theoretical results, since the Rao-Blackwellization Step \ref{step:2 bias correction} removes instrument selection bias. Moreover, we observe that the standard deviation of the RIVW estimator is not sensitive to $\eta$ either, as the five curves in panel (B) corresponding to different values of $\eta$ almost coincide. We fix $\eta=0.5$ as it leads to satisfactory performance for the RIVW estimator across a range of simulation settings
    
%--------------------------------------------------------------------------------
% Simulations setup and results
%--------------------------------------------------------------------------------
\subsection{Simulations setup and results}\label{Section:simulation-no-pleiotropy}
In this section, we compare the performance of our RIVW estimator with the classical two-/three-sample IVW, the debiased IVW (dIVW) \citep{ye2019debiased}, and the two-/three-sample RAPS estimator \citep{zhao2020statistical} in simulation studies. We generate $2,000$ Monte Carlo samples with the true causal effect $\beta=0.2$ and $p = 200,000$ independent SNPs. \xwm{The true SNP-exposure associations, $\gamma_j$ follow the mixture distribution:
\begin{align*}
\begin{pmatrix}
\gamma_j \\[5pt]
\alpha_j
\end{pmatrix} &\sim \pi_{x}\rho \begin{pmatrix}
\mathcal{N}(0,\varepsilon_x^2)\\[5pt]
\delta_0
\end{pmatrix}   + \pi_{x}(1-\rho) \begin{pmatrix}
\mathcal{N}(0,\varepsilon_x^2)\\[5pt]
\mathcal{N}(0,\tau^2)
\end{pmatrix} + 
\pi_{y} \begin{pmatrix}
\delta_0\\[5pt]
\mathcal{N}(0,\tau^2)
\end{pmatrix} +
(1-\pi_x - \pi_y) \begin{pmatrix}
\delta_0\\[5pt]
\delta_0
\end{pmatrix},
\end{align*}
and the true SNP-outcome associations $\Gamma_j = \beta\gamma_j + \alpha_j$.
Here, $\alpha_j$ captures balanced horizontal pleiotropy (see Assumption \ref{assump:Measurement-error-pleiotropy}$^\prime$), $\mathcal{N}(0,\varepsilon^2)$ represents a normal distribution with mean $0$ and variance $\varepsilon^2$, and $\delta_0$ is the Dirac measure centred at zero. The first two components of the above mixture distribution correspond to relevant IVs (i.e., having nonzero associations with the exposure), and $1-\rho$ controls the fraction of relevant IVs that also exhibit pleiotropic effects. The third term represents IVs that are only associated with the outcome but not the exposure, while the last component in the mixture distribution corresponds to SNPs that have no association with either the outcome or the exposure.} Similar data generating models have been widely used in the literature \citep{qi2019mendelian,zeng2018signatures}. For future reference, we define heritability as the proportion of the phenotypic variance explained by additive effects of all genetic variants \citep{yang2017concepts}. Because the data have been standardized and the SNPs are assumed to be independent, the heritability of exposure ($h_x^2$) and outcome ($h_y^2$) can be calculated as
\begin{align*}
h_x^2 &= \Var\Big[\sum_{j=1}^{p}\gamma_j G_j\Big] \ /\ \Var[X] = \sum_{j=1}^{p}\gamma_j^2\Var[G_j] \approx p \pi_x \varepsilon_x^2,\\
h_y^2 &= \Var\Big[\sum_{j=1}^{p}\Gamma_j G_j\Big]\ /\ \Var[Y] = \sum_{j=1}^{p}\Gamma_j^2 \Var[G_j] \approx \beta^2 h_x^2 + p\left(\pi_x(1-\rho) + \pi_y\right) \tau^2.
\end{align*}
For simplicity, we set $\rho=1$, $\pi_x=\pi_y$, and $\varepsilon_x^2 = \tau^2$ to conserve space. Additional simulation evidence is available in the Supplementary Material \citep{Ma-Wang-Wu-2022-AOS-Supp}. The parameter values used to generate Monte Carlo samples are given in corresponding tables. 
	
Following common practices, the cutoff value $\lambda$ is either $5.45$ (corresponding to the significance threshold $5\times 10^{-8}$) for the two-sample IVW estimators, or $0$ (using all available SNPs) for the dIVW estimator. As three-sample MR involves a third independent sample for IV selection and are immune to winner's curse bias, we also consider the cutoff value $\lambda=4.06$ (corresponding to the significant threshold $5\times 10^{-5}$). Note that in three-sample MR analyses, we generate a new independent exposure data with the same sample size for IV selection. 
	
We report our simulation results with six measures: ``$\hat{\beta}$'' (average causal effect estimate across Monte Carlo samples), ``monte SD'' (Monte Carlo standard deviation), ``SE'' (average standard error across Monte Carlo samples), ``CP'' (average coverage probability of the 95\% confidence interval), ``length'' (average length of the 95\% confidence interval), and ``\# IVs'' (number of selected IVs, averaged across Monte Carlo samples). Table~\ref{Table:simRes} summarizes the performance of various estimators under different settings, which we discuss below. 
	
For the two-sample IVW estimator, our simulation results confirm that it is biased towards zero due to winner's curse bias and measurement error bias, resulting in confidence intervals that under cover the true causal effect. In the Supplementary Material (Table S.8 of \citep{Ma-Wang-Wu-2022-AOS-Supp}), we further demonstrate that the two-sample IVW estimator with a liberal cutoff value ($\lambda = 4.06$) yields a larger bias than that with a stringent cutoff ($\lambda = 5.45$). 
	
For the two-sample dIVW estimator, we have followed the recommendation of \cite{ye2019debiased} and used all available IVs (i.e., $\lambda = 0$) to estimate $\beta$. While dIVW indeed provides accurate causal effect estimates when the average IV strength is relatively high (medium and high heritability cases), it exhibits comparably high variance among the considered estimators in the presence of too many weak instruments (for example, in the low heritability situation).
	
For the three-sample IVW estimator, we observe that it tends to produce slightly biased causal effect estimates due to its failure to account for measurement error bias. By correcting measurement error bias, both the three-sample dIVW estimator and three-sample RAPS estimator yield accurate causal effect estimates. Furthermore,  three-sample RAPS with a liberal cutoff values ($\lambda = 4.06$) yields narrower confidence intervals than that with a stringent cutoff value ($\lambda = 5.45$). \xwm{This indicates that a more liberal cutoff can be more preferable when both winner's curse bias and measurement error bias are fully removed.}

By fully correcting measurement error bias and winner's curse bias, our RIVW estimator provides accurate causal effect estimates, and the resulting confidence intervals have coverage probabilities close to the nominal 95\% level. Furthermore, because the RIVW estimator is able to accommodate a liberal cutoff value ($\lambda = 4.06$) and incorporate more IVs to estimate the causal effect, its confidence intervals have similar lengths to those produced by  three-sample dIVW/RAPS. We also note that standard errors provided by our formula are  close to the Monte Carlo standard deviations, validating our theoretical investigations (Theorem \ref{thm:valid standard error}). Finally, the sRIVW estimator (proposed in Section \ref{section:smoothing-over-randomization}) that smooths out the effect of IV selection yields similar results. \xwm{We do not report the number of selected instruments for the sRIVW estimator, as it employs a re-weighting strategy instead of IV selection.}
 
\begin{table}[!tbhp]
\centering
\renewcommand{\arraystretch}{1.2}
\begin{tabular}{lcccccc}
\hline\hline
& $\hat{\beta}$ & monte SD & SE & CP & length & \# IVs \\ 
\hline
\multicolumn{7}{c}{Low heritability: \xwm{$\pi_x=\pi_y=0.002,\varepsilon_x^2=\tau^2 = 1\times  10^{-4}, h_x^2 = 0.040, h_y^2 = 0.042$}}  \\ 
  IVW ($\lambda=5.45$) & 0.182 & 0.023 & 0.023 & 0.865 & 0.090 & 40 \\ 
  dIVW ($\lambda = 0$) & 0.209 & 0.125 & 0.123 & 0.959 & 0.484 & 200000 \\ 
  RIVW ($\lambda = 4.06,\eta =0.5$) & 0.200 & 0.022 & 0.022 & 0.951 & 0.087 & 148 \\ 
  sRIVW ($\lambda = 4.06,\eta =0.5$) & 0.200 & 0.021 & 0.021 & 0.947 & 0.082 & \xwm{---} \\ 
  Three-sample IVW ($\lambda=5.45$) & 0.196 & 0.025 & 0.025 & 0.950 & 0.097 & 40 \\ 
  Three-sample IVW ($\lambda=4.06$) & 0.200 & 0.020 & 0.021 & 0.954 & 0.081 & 99 \\ 
  Three-sample RAPS ($\lambda=5.45$) & 0.200 & 0.026 & 0.028 & 0.962 & 0.108 & 40 \\ 
  Three-sample RAPS ($\lambda=4.06$) & 0.200 & 0.021 & 0.022 & 0.958 & 0.085 & 99 \\ 
  Three-sample dIVW ($\lambda=5.45$) & 0.201 & 0.026 & 0.026 & 0.954 & 0.101 & 40 \\ 
  Three-sample dIVW ($\lambda=4.06$) & 0.200 & 0.020 & 0.021 & 0.954 & 0.081 & 99 \\ 
   \hline
\multicolumn{7}{c}{Medium heritability: \xwm{$\pi_x=\pi_y=0.01,\varepsilon_x^2=\tau^2 = 1\times  10^{-4}, h_x^2 = 0.200, h_y^2 = 0.208$}} \\ 
IVW ($\lambda=5.45$) & 0.182 & 0.011 & 0.010 & 0.566 & 0.040 & 200 \\ 
  dIVW ($\lambda = 0$) & 0.201 & 0.025 & 0.024 & 0.941 & 0.096 & 200000 \\ 
  RIVW ($\lambda = 4.06,\eta =0.5$) & 0.200 & 0.010 & 0.009 & 0.944 & 0.037 & 509 \\ 
  sRIVW ($\lambda = 4.06,\eta =0.5$) & 0.200 & 0.009 & 0.009 & 0.947 & 0.036 & \xwm{---} \\ 
  Three-sample IVW ($\lambda=5.45$) & 0.195 & 0.011 & 0.011 & 0.917 & 0.043 & 200 \\ 
  Three-sample IVW ($\lambda=4.06$) & 0.200 & 0.009 & 0.009 & 0.956 & 0.036 & 452 \\ 
  Three-sample RAPS ($\lambda=5.45$) & 0.199 & 0.012 & 0.012 & 0.951 & 0.047 & 200 \\ 
  Three-sample RAPS ($\lambda=4.06$) & 0.200 & 0.009 & 0.010 & 0.955 & 0.037 & 452 \\ 
  Three-sample dIVW ($\lambda=5.45$) & 0.200 & 0.011 & 0.011 & 0.947 & 0.045 & 200 \\
    Three-sample dIVW ($\lambda=4.06$) & 0.200 & 0.009 & 0.009 & 0.956 & 0.036 & 452 \\ 

   \hline
\multicolumn{7}{c}{High heritability: \xwm{$\pi_x=\pi_y=0.01,\varepsilon_x^2=\tau^2 = 3\times  10^{-4}, h_x^2 = 0.600, h_y^2 = 0.624$}} \\ 
  IVW ($\lambda=5.45$) & 0.193 & 0.005 & 0.004 & 0.688 & 0.017 & 655 \\ 
  dIVW ($\lambda = 0$) & 0.200 & 0.010 & 0.009 & 0.928 & 0.035 & 200000 \\ 
  RIVW ($\lambda = 4.06,\eta =0.5$) & 0.200 & 0.005 & 0.004 & 0.952 & 0.018 & 993 \\ 
  sRIVW ($\lambda = 4.06,\eta =0.5$) & 0.200 & 0.005 & 0.004 & 0.952 & 0.017 & \xwm{---} \\ 
  Three-sample IVW ($\lambda=5.45$) & 0.197 & 0.005 & 0.005 & 0.908 & 0.018 & 655 \\ 
  Three-sample IVW ($\lambda=4.06$) & 0.200 & 0.004 & 0.004 & 0.955 & 0.017 & 943 \\ 
  Three-sample RAPS ($\lambda=5.45$) & 0.200 & 0.005 & 0.005 & 0.959 & 0.019 & 655 \\ 
  Three-sample RAPS ($\lambda=4.06$) & 0.200 & 0.005 & 0.005 & 0.954 & 0.018 & 943 \\ 
  Three-sample dIVW ($\lambda=5.45$) & 0.200 & 0.005 & 0.005 & 0.960 & 0.018 & 655 \\ 
  Three-sample dIVW ($\lambda=4.06$) & 0.200 & 0.004 & 0.004 & 0.955 & 0.017 & 943 \\ 
   \hline\hline
\end{tabular}
\caption{Simulation results under different settings. We set the true effect $\beta = 0.2$. The columns are: estimated effect size ($\hat{\beta}$), Monte Carlo standard deviation (monte SD), average standard error (SE), empirical coverage probability of 95\% confidence intervals (CP), average confidence interval length (length), and number of selected instruments (\# IVs). $h_x^2$ and $h_y^2$ are the heritability of exposure and outcome.  }
\label{Table:simRes}
\end{table}

We provide additional simulation evidence in the Supplementary Material \citep{Ma-Wang-Wu-2022-AOS-Supp} for a wider range of model specifications, such as different sample sizes for the exposure and outcome GWAS data (Tables S.5 and S.6), varying the sample size for each SNP (Table S.7), and $\beta=0$ (i.e., no causal effect between the exposure and the outcome, Table S.9). In summary, the simulation results confirm that our RIVW estimator continues to deliver accurate causal effect estimation and valid statistical inference.

%%%%%%%%%%%%%%%%%%%%%%%%%%%%%%%%%%%%%%%%%%%%%%%%%%%%%%%%%%%%%%%%%%%%%%%%%%%%%%%%%
% Real data applications
%%%%%%%%%%%%%%%%%%%%%%%%%%%%%%%%%%%%%%%%%%%%%%%%%%%%%%%%%%%%%%%%%%%%%%%%%%%%%%%%% 
\section{Real data applications}\label{Sec:real-data}
	
We conduct two real data analyses. First, the same-trait type analysis \citep{zhao2020statistical} demonstrates that the performance of the proposed RIVW estimator is in line with our simulations and theory. Second, the study of the relationship between BMI and COVID-19 severity provides new perspectives on the benefit of body weight management. 

%--------------------------------------------------------------------------------
% Data harmonization
%--------------------------------------------------------------------------------
\subsection{Data harmonization}\label{Sec:extension-pruning}
	
We harmonize the data through the following steps. First, genetic variants that are not available in the outcome dataset are excluded. Second, we select independent genetic variants that have no linkage disequilibrium (R-Squared $ < 0.001$ with an extension of 10,000 Kb in the genome, which is suggested by the \texttt{twosampleMR} package \cite[version 0.5.5;][]{hemani2018mr} and has been widely adopted in applied MR studies) with other selected genetic variants through the revised sigma-based pruning procedure introduced in Section \ref{Sec:extension-correlated-IV}. Specifically, when genetic variants are in linkage disequilibrium, we select the variant with the smallest  standard deviation of the SNP-exposure association. We employ this revised sigma-based pruning procedure because standard clumping, which is based on p-value comparisons, introduces a different type of selection bias; see \cite{robertson2016accounting} for related discussion, and Tables S.1 and S.2 in the Supplementary Material \citep{Ma-Wang-Wu-2022-AOS-Supp}. Third, by leveraging allele frequency information, we infer the strand direction of ambiguous SNPs and  harmonize exposure-outcome datasets by using the \texttt{twosampleMR} package. We use the default setting with $\lambda=4.06$ and $\eta = 0.5$ for our proposed RIVW estimator, and set $\lambda = 0$ and $\lambda = 5.45$ for the dIVW and IVW estimators, respectively.
	
%--------------------------------------------------------------------------------
% Same trait analyses
%--------------------------------------------------------------------------------
\subsection{Same trait analyses}\label{Sec:sametrait}
	
To investigate the consequences of measurement error bias and winner's curse bias in real data applications, we use two separate GWAS datasets on the same trait as the exposure and outcome. The true causal effect is 1, and therefore such analyses provide opportunities to validate our proposed method with real world data. We conduct two same trait analyses: body mass index (BMI)--BMI analysis and high-density lipoprotein (HDL)--HDL analysis. In the BMI--BMI analysis, we use two BMI GWAS results from the UK Biobank  \citep[sample size 461,460, ID: ukb-b-19953;][]{bycroft2018uk} and from the GIANT consortium \citep[sample size 234,069, ID: ieu-a-2;][]{locke2015genetic}, denoted by BMI-1 and BMI-2, respectively. In the HDL--HDL analysis, we use two HDL GWAS results from the UK Biobank \citep[sample size 403,943, ID: ukb-b-109;][]{bycroft2018uk} and the Global Lipids Genetics Consortium \citep[sample size 94,595, ID: ebi-a-GCST002223;][]{willer2013discovery}, denoted by HLD-1 and HDL-2, respectively.
	
\begin{table}[ht]
\renewcommand{\arraystretch}{1.1}
\renewcommand{\arraystretch}{1.2}
\centering
\begin{tabular}{lccccc|ccccc}
\hline\hline
& $\hat{\beta}$ & SE & 95\% CI & \# IVs & F & $\hat{\beta}$ & SE & 95\% CI & \# IVs & F\\ 
\hline
&\multicolumn{5}{c|}{BMI-1--BMI-2 analysis}&	\multicolumn{5}{c}{HDL-1--HDL-2 analysis}\\
RIVW &  1.030 & 0.024 & 0.983-1.077 & 913 & 21.9 & 1.022 & 0.036 & 0.952-1.092 & 766 & 26.8 \\ 
dIVW & 0.981 & 0.034 & 0.915-1.048 & 1811 & 3.7 & 0.912 & 0.048 & 0.819-1.006 & 1787 & 3.6 \\ 
IVW & 0.785 & 0.015 & 0.756-0.814 & 435 & 42.5 & 0.915 & 0.020 & 0.876-0.953 & 321 & 68.1\\ 
\hline
&\multicolumn{5}{c|}{BMI-2--BMI-1 analysis}&	\multicolumn{5}{c}{HDL-2--HDL-1 analysis}\\
RIVW & 1.194 & 0.060 & 1.076-1.312 & 401 & 18.1  & 1.039 & 0.048 & 0.944-1.134 & 380 & 20.9  \\ 
dIVW &  1.402 & 0.070 & 1.266-1.539 & 1810 & 2.1 &0.887 & 0.050 & 0.789-0.984 & 1792 & 2.1 \\ 
IVW &  0.870 & 0.026 & 0.820-0.920 & 82 & 42.0 & 0.922 & 0.022 & 0.878-0.967 & 99 & 58.1  \\ 
 \hline\hline
\end{tabular}
\caption{Same trait analyses results. The true causal effect is $\beta = 1$. $\hat{\beta}$, SE,  95\% CI, \# IVs, and F stand for the estimated effect size, its standard error, the 95\% confidence interval,  the number of selected IVs, and the F statistic, respectively. }
\label{realdata1}
\end{table}
	
The results are summarized in Table~\ref{realdata1}. Overall, the RIVW estimator provides accurate estimates of the causal effect $\beta = 1$, confirming its ability to remove both measurement error bias and winner's curse bias in the two-sample MR framework. In addition, RIVW yields smaller standard errors compared with the dIVW estimator, indicating that instrument selection may help improve estimation efficiency. The IVW estimator is biased downwards due to the two sources of biases. While the dIVW estimator (with $\lambda = 0$) generally performs well, we notice that the corresponding F statistic (average of $(\hat{\gamma}_j/\sigma_{X_j})^2$ across instruments) is quite small, suggesting that many of the included IVs are not relevant for estimating the causal effect. We also observe that reversing the exposure and outcome GWAS leads to quite different numbers of instruments being selected. This is because the UK Biobank study (BMI-1 and HDL-1) has a much larger sample size. As a result, associations calculated from the UK Biobank study tend to have smaller standard errors on average. This should explain why more IVs are selected in the BMI-1--BMI-2 and HDL-1--HDL-2 analyses. Because both BMI-2--BMI-1 and HDL-2--HDL-1 analyses employ fewer IVs and the F statistics also become smaller, our RIVW estimator is more likely to produce noisy estimates in finite samples. Finally, compared with Table S.3 in the Supplementary Material \citep{Ma-Wang-Wu-2022-AOS-Supp} which employs standard pruning (random instrument selection without the use of any GWAS summary statistic), we note that the revised sigma-based pruning procedure leads to significant efficiency improvement. 

%--------------------------------------------------------------------------------
% BMI and COVID-19 severity
%--------------------------------------------------------------------------------
\subsection{BMI and COVID-19 severity}\label{Sec:covid}
Epidemiological studies have reported associations between BMI and COVID-19 illness \citep{soeroto2020effect}, and several MR analyses have confirmed that higher BMI is a causal risk factor (exposure) for COVID-19 severity \citep{leong2021cardiometabolic,ponsford2020cardiometabolic}. However, the causal effect estimated from these MR analyses may be biased due to  measurement error bias and winner's curse bias. While identifying  causal risk factors is important for the general public to combat COVID-19, providing accurate causal effect estimates is also important, as it both facilitates effective communication and offers solid evidence for public health policy-making. We apply our proposed method with BMI from the UK Biobank (sample size 461,460, ID: ukb-b-19953) as the exposure and COVID-19 severity from the covid-19hg \citep[v6 leave out UK Biobank samples;][]{ganna2021mapping} as the outcome data. Detailed information on participating studies, quality control, and analyses have been provided on the COVID-19 HGI website (http://www.covid19hg.org/results/). In brief, data from 20,980 hospitalized COVID-19 patients and 1,628,780 population controls were used. The hospitalized COVID-19 cases represent patients with (i) laboratory confirmed SARS-CoV-2 infection (RNA and/or serology based) and (ii) hospitalization due to COVID-related symptoms. 
	
Table~\ref{realdat2} summarizes our results. First, we confirm that BMI is a risk factor for COVID-19 severity as all three methods yield statistically significant estimates. Similar to the same trait analysis, our RIVW approach provides smaller standard errors compared with the dIVW estimator. Because the proposed RIVW estimator lifts the winner's curse and removes measurement error bias, it produces a larger estimate. This larger estimate indicates that BMI is a critical risk factor (maybe even more important than we originally anticipated), and effectively communicating this finding may encourage the general public to take extra precautions and reduce COVID-19 related hospitalization. 

\begin{table}[ht]
\renewcommand{\arraystretch}{1.1}
\centering
\begin{tabular}{rccccc}
\hline  \hline
& $\hat{\beta}$ & SE & 95\% CI & \# IVs & F \\ 
\hline
RIVW & 1.60 & 0.142 & 1.32-1.88 & 1025 & 21.4 \\   dIVW & 1.49 & 0.163 & 1.17-1.81 & 1985 & 4.0 \\ 
 IVW & 1.25 & 0.069 & 1.11-1.39 & 477 & 42.4 \\ 
\hline
\end{tabular}
\caption{\xwm{BMI and COVID-19 severity. $\hat{\beta}$, SE,  95\% CI, \# IVs, and F stand for the estimated effect size (odds ratio), its standard error, the 95\% confidence interval, the number of selected IVs, and the F statistic, respectively.}}
\label{realdat2}
\end{table}

%%%%%%%%%%%%%%%%%%%%%%%%%%%%%%%%%%%%%%%%%%%%%%%%%%%%%%%%%%%%%%%%%%%%%%%%%%%%%%%%%
% Closing remarks
%%%%%%%%%%%%%%%%%%%%%%%%%%%%%%%%%%%%%%%%%%%%%%%%%%%%%%%%%%%%%%%%%%%%%%%%%%%%%%%%%
\section{Closing remarks}
	
We introduced a new two-sample Mendelian randomization with summary data framework that breaks the winner's curse. We further propose the rerandomized inverse variance weighted estimator that eliminates both winner's curse bias and measurement error bias in the popular IVW estimator. To facilitate statistical hypothesis testing, we also provide valid standard error construction. Our RIVW estimator is easy to implement and enjoys rigorous theoretical guarantees. It also remains valid in the presence of balanced horizontal pleiotropy. We plan to further study winner's curse removal in the presence of correlated instuments, and extend our framework to robust MR analyses. 

%\section{Funding}
\ \bigskip \\
\noindent \textbf{Funding.} The authors acknowledge the support of the National Science Foundation (DMS-2015325) and the National Institute of Health (1R03AG070669, R01MH125746). The content is solely the responsibility of the authors and does not necessarily represent the official views of the National Institutes of Health.

\bibliographystyle{jasa} 
	\bibliography{reference}
\end{document}